\documentclass[twocolumn,english,prd,nofootinbib,final,showpacs]{revtex4}
\usepackage[T1]{fontenc}
\usepackage[latin1]{inputenc}
\usepackage{graphicx}

\makeatletter

\providecommand{\tabularnewline}{\\}

\usepackage{graphicx}

\usepackage{graphicx}

\font \bolditalics = cmmib10

\newcommand{\boldmm}[1]{\textfont1=\bolditalics \hbox{$\bf#1$}}

\newcommand{\gammag}{\boldmm{\gamma}}
\newcommand{\thetag}{\boldmm{\theta}}

\newcommand{\Sg}{{\bf S}}
\newcommand{\dg}{{\bf d}}
\newcommand{\sg}{{\bf s}}
\newcommand{\OmM}{\Omega_0}

\usepackage{babel}
\makeatother
\begin{document}

\title{Constraining Dark Energy Evolution with Gravitational Lensing by
Large Scale Structures}

\author{Karim Benabed}

\affiliation{Center for Cosmology and Particle Physics, Physics Department, New
York University, New York, NY 10003,USA}

\email{karim.benabed@physics.nyu.edu}

\author{Ludovic Van Waerbeke}

\affiliation{Institut d'Astrophysique de Paris, 98 bis Bd Arago, 75014 Paris,France}

\email{waerbeke@iap.fr}

\begin{abstract}
We study the sensitivity of weak lensing by large scale structures
to the evolution of dark energy. We explore a 2-parameters model of
dark energy evolution, inspired by tracking quintessence models. To
this end, we compute the likelihood of a few representative models
with varying and non varying equation of states. Based on an earlier
work, we show that the evolution of dark
energy has a much stronger impact on the non-linear structure growth
than on the angular diameter distance, which makes large scale structure
measurements a very efficient probe of this evolution. For the different
models, we investigate the dark energy parameters degeneracies with
the mass power spectrum shape $\Gamma$, normalisation $\sigma_{8}$,
and with the matter mean density $\Omega_{M}$. This result is a strong
motivation for performing large scale structure simulations beyond
the simple constant dark energy models, in order to calibrate the
non-linear regime accurately. Prospective for the Canada France Hawaii
Telescope Legacy Survey (CFHTLS) and Super-Novae Acceleration Probe
(SNAP) are given. These results complement nicely the  cosmic microwave
background and Super-Novae constraints. Weak lensing is shown to be
more sensitive to a variation of the equation of state, whereas CMB and
SNIa give information on its constant part. 
\end{abstract}


\maketitle
Dark energy is a generic way to describe the acceleration of the universe.
Within this framework, the acceleration of the expansion measured
by the type Ia Super-Novae surveys \cite{1999ApJ...517..565P,2001ApJ...560...49R}
is explained by the contribution of a new component. In a FRLW metric,
this component is described by its equation of state. In this paradigm,
the cosmological constant is one possible model, among others, of
dark energy. 

The question of the properties of this dark energy remains open. Different
observations have been proposed to evaluate them. Measurement of the
distances to Super-Novae \cite{2001PhLB..500....8A,2001PhRvL..86....6M,2001AA...380....6G}
or of the size of structures on the CMB \cite{2000PhRvD..62j3505B,2002MNRAS.330..965D}
provide informations on how the dark energy modifies the relation
between cosmological distances and redshifts. The evolution of large
scale structures also probe the properties of dark energy. This kind
of tests can be built using cluster abundances \cite{2001ApJ...553..545H,2002PhRvL..88w1301W,2003PhRvD..67h1304H},
Ly-$\alpha$ forest \cite{2002astro.ph.12343S}, strong \cite{2002astro.ph.10066B,2002AA...393..757S}
and weak lensing\cite{BB01,2002PhRvD..65f3001H,2002PhRvD..66h3515H,2002AA...396...21B,2003MNRAS.341..251W}... 

In this article, we will investigate the weak lensing constraints
in the case of a varying equation of state. This case has been first
investigated qualitatively by Benabed and Bernardeau \cite{BB01}
(hereafter BB01). We will expand their results and propose the first
quantitative analysis of the efficiency of the shear two points function
to probe varying equation of state dark energy models. The constraints
on the dark energy using the linear regime alone are not particularly
strong, even using the tomography technique \cite{2002PhRvD..66h3515H}.
Here, we take into account the non-linear regime of gravitational
collapse, which is known to convey much of the dark energy sensitivity
\cite{BB01}, and we study the degeneracy of the dark energy parameters
with the matter density $\Omega_{M}$, the mass power spectrum shape
$\Gamma$ and normalisation $\sigma_{8}$. 

As said above, large scale structures are sensitive to the expansion
rate of the universe, which makes the structure growth sensitive to
the dark energy content of the Universe \cite{BB01}. This sensitivity
is due to the fact that when the dark energy starts to dominate the
energy budget of the Universe, the efficiency of the gravitational
collapse is reduced. Hence, the density fluctuations growth changes
when the dark energy differs from a simple cosmological constant.
For models within the current SNIa constraints, it corresponds to
a slower growth in the linear regime. The stronger the variation in
the evolution of dark energy, the earlier this effect occurs and the
more suppressed will be the structure growth. Keeping the amplitude
of density fluctuations fixed today, this translates into an earlier
entrance of the fluctuations in the non-linear regime, leading to
more concentrated dark matter halos\cite{2003astro.ph..3304K,2003astro.ph..5286L},
and therefore stronger lenses \cite{2002AA...393..757S}. As shown
in BB01, the shear two points function is sensitive to these two effects.
It provides an unbiased measure of the projected density power spectrum
in both the linear and non-linear regimes, which is a direct test
of the evolution of large scale structures, and through it, a measure
of the properties of dark energy. In particular, the scale at which
the transition between the two regimes occurs is tested. This feature
is the key to constraint the properties of dark energy. 

In the following, we first review the computation of the shear two
points function with a dark energy component. Then, we propose a simple
two parameters model that encompass the major features of a generic
class of dark energy, namely the tracking quintessence models. Next,
we address the question of the efficiency of future lensing surveys
to determine the dark energy properties.

\section{Theory\label{sec:Model-building}}

We establish in this section the theoretical basis of the result presented
in section \ref{sec:results}. We show here how to compute the non-linear
cosmic shear power spectrum with a non-trivial dark energy, and how
it is evaluated from the data. We discuss the sensitivity of this
quantity to the evolution of an hypothetical dark energy component.
We then propose a dark energy equation of state parameterization suitable
for a class of dark energy models.

\subsection{Cosmology\label{sub:Cosmology}}

Lets assume that the dark energy component is minimally coupled to
the universe, which means that it interacts with the rest of the Universe
via gravity only. The expansion of the universe is completely described
by the Friedmann equations\begin{eqnarray}
\left(\frac{\dot{a}}{a}\right)^{2} & = & \frac{8\pi}{3M_{\textrm{Planck}}}\sum\rho_{\textrm{X}}\label{eq:fried1}\\
\frac{\ddot{a}}{a} & = & -\frac{4\pi}{3M_{\textrm{Planck}}}\sum\left(\rho_{\textrm{X}}+3p_{\textrm{X}}\right).\label{eq:fried2}\end{eqnarray}
 and the knowledge of an equation of state for each component\begin{equation}
P_{\textrm{X}}=w_{\textrm{X}}\,\rho_{\textrm{X}}.\end{equation}
 The radiation, matter and curvature equation of state are fixed.
The only unknown quantity here, is the equation of state parameter
of the dark energy, $w_{Q}$. It usually varies between $+1$ and
$-1$. The case $w_{Q}=-1$ corresponds to a cosmological constant. 

It has been proposed recently that $w_{Q}$ can also take values lower
than $-1$ \cite{2003astro.ph..1273C}. This is only possible if the
dark energy has a negative kinetic energy. Such unusual behavior have
only been found in very specific models\cite{Onemli:2002hr}, and
it is not yet clear if it has any physical meaning. Therefore, we
keep the conservative prior $w_{Q}\geq-1$. 

We also assume that the dark energy does not fluctuate and thus is
not coupled to the fluctuations of the matter density. This is of
course not true in general: it is expected that dark energy fluctuates
from one Hubble volume to the next, which is expected to leave an
imprint on the metric fluctuations when the universe expands. However,
it has been showed that these fluctuations are quickly damped by the
evolution. Moreover, it is expected that they are negligible on scales
smaller than the horizon at recombination \cite{1998PhRvD..58b3503F,2002PhRvD..66l3506M,2000PhRvD..62j3505B}.
In the following, we only consider such scales and can thus safely
ignore these fluctuations. 

With these assumptions, the impact of dark energy on the evolution
of the large structures is completely described by the equation of
state \cite{BB01}.

\subsection{Shear measurements on distant galaxies}

The deviation of light by the gravitational potential wells distorts
the image of the distant galaxies. This shear effect can be used to
probe the projected mass distribution along the line of-sight (see
and references \cite{Bartelmann:1999yn} therein), from a measurement
of the shape of the lensed galaxies. The lensing effect produced by
the large scale structures is weak, but has already been measured
\cite{2003astro.ph..5089V}. 

The gravitational lensing effect depends on the second order derivatives
of the gravitational potential projected along the line-of-sight.
The convergence $\kappa$ and the shear $\gammag$ describe the distortion
of the image of the distant images (located at some redshift $z_{s}$),
by the inhomogeneous matter distribution along the line-of-sight.
At linear order, convergence and shear field are related \begin{equation}
\Delta\kappa=\left(\partial_{1}^{2}-\partial_{2}^{2}\right)\gamma_{1}+2\partial_{1}\partial_{2}\gamma_{2}\label{eq:kappagamma}\end{equation}
 At the same order, the convergence in the direction $\thetag$, which
describes the isotropic change of the image at position $\chi_{s}$,
is given by: 

\begin{eqnarray}
\kappa(\thetag) & = & {\frac{3}{2}}{\frac{H_{0}}{c}}^{2}\OmM\label{eq:kappaeq}\\
 &  & \hspace{-1cm}\int_{0}^{\chi_{S}}{\textrm{d}}\chi\;{\frac{\mathcal{{D}}(\chi_{s}-\chi)\,\mathcal{{D}}(\chi)}{\mathcal{{D}}(\chi_{s})}}{\frac{\delta\left(\mathcal{{D}}(\chi)\thetag,\chi\right)}{a(\chi)}},\nonumber \end{eqnarray}
 where $\chi_{s}(z_{s})$ is the source radial distance located at
redshift $z_{s}$, and $a=1/(1+z)$ is the scale factor. The radial
distance at redshift $z$ is given by 

\begin{equation}
\chi(z)=\int_{0}^{z}{\textrm{d}}z'{\frac{c}{H}}.\label{eq:cosmo:dist}\end{equation}
 The angular diameter distance $\mathcal{{D}}$ is defined by \begin{equation}
\mathcal{{D}}(\chi)=\left\{ \begin{array}{ll}
\sin(\sqrt{K}\chi)/\sqrt{K}, & K>1\\
\chi, & K=1\\
\sinh(\sqrt{K}\chi)/\sqrt{K}, & K<1\end{array}\right.\end{equation}
 where $K$ is the curvature. As a consequence, the weak lensing effect
is a direct and unbiased measurement of the projected density contrast. 

We focus on the convergence power spectrum $P_{\kappa}(\ell)$. It
can be shown \cite{1997AA...322....1B,1997ApJ...484..560J} that it
is directly related to the 3-dimensional mass power spectrum $P_{3D}$
via: 

\begin{eqnarray}
P_{\kappa}(\ell) & = & {\frac{9}{4}}\left(\frac{H_{0}}{c}\right)^{4}\OmM^{2}\label{eq:pofkappa}\\
 &  & \hspace{-1cm}\int_{0}^{\chi_{S}}{\textrm{d}}\chi'\;{\frac{g^{2}(\chi')}{\mathcal{D}^{2}(\chi')a^{2}(\chi')}}P_{3D}\left({\frac{\ell}{\mathcal{D}(\chi')}},\chi'\right),\nonumber \end{eqnarray}
 where the $g(\chi)$ function describe the lensing efficiency, \begin{equation}
g(\chi)=\frac{\mathcal{{D}}(\chi_{s}-\chi)\,\mathcal{{D}}(\chi)}{\mathcal{{D}}(\chi_{s})}.\label{eq:gfunc}\end{equation}
 When the lensed galaxies are distributed in redshift, the observed
signal is given by Eq.\ref{eq:gfunc} integrated among the source
redshift distribution $p_{s}(\chi)$. In that case, the $g(\chi)$
function becomes: 

\begin{equation}
g(\chi)=\mathcal{D}(\chi)\int_{\chi}^{\chi_{s}}{\textrm{d}}\chi'\; p_{s}(\chi')\,{\frac{\mathcal{D}(\chi_{s}-\chi')}{\mathcal{D}(\chi_{s})}}.\label{eq:gfunc:pz}\end{equation}
 The source redshift distribution $p_{s}(z)$ is normalised, and usually
parametrised as \begin{equation}
p_{s}(z)=\Gamma^{-1}\left({\frac{1+\alpha}{\beta}}\right){\frac{\beta}{z_{s}}}\left({\frac{z}{z_{s}}}\right)^{\alpha}~\exp\left[-\left({\frac{z}{z_{s}}}\right)^{\beta}\right].\label{eq:pz:def}\end{equation}
 The free parameters $\alpha$, $\beta$ and $z_{s}$ are adjusted
to accommodate different survey properties. 

The ellipticity of the galaxies is an unbiased measure of the shear,
from which we derive the statistical properties of the convergence
field (see a review of the observational results in \cite{2003astro.ph..5089V}).
In practice, the statistic of interest is the aperture mass variance
as function of scale, $\langle M_{\textrm{ap}}^{2}(\theta)\rangle$
\cite{1998MNRAS.296..873S}, also called the $M_{\textrm{ap}}$ statistic.
It links the variance of the convergence field (which is the field
of physical interest, because proportional to the projected mass density)
and the shear (which is the observable quantity). Its main feature
is to provide a natural separation between the cosmological signal
(which is curl-free) and the systematics (contributing to the curl
mode). It has already been measured on several galaxy surveys \cite{2003astro.ph..5089V}.
The $M_{\textrm{ap}}$ statistic at a scale $\theta_{c}$ is defined
as the convergence smoothed with a compensated filter $U(\thetag)$.
Using Eq. (\ref{eq:kappagamma}), it is also given by a properly smoothed
shear component $\gamma_{t}$: 

\begin{eqnarray}
M_{\textrm{ap}} & = & \int_{0}^{\theta_{c}}{\textrm{d}}^{2}\theta\; U(\theta)\;\kappa(\thetag)\\
 & = & \int_{0}^{\theta_{c}}{\textrm{d}}^{2}\theta\; Q(\theta)\;\gamma_{t}(\thetag),\end{eqnarray}
 with $\int_{0}^{\theta_{c}}{\textrm{d}}\theta\;\theta\; U(\theta)=0$,
and where\begin{equation}
Q(\theta)={\frac{2}{\theta_{c}^{2}}}\int_{0}^{\theta_{c}}{\textrm{d}}\theta'\;\theta'\, U(\theta')-U(\theta).\end{equation}
 The tangential shear $\gamma_{t}$ at a location $\thetag=(\theta\cos\varphi,\theta\sin\varphi)$
is defined by 

\begin{equation}
\gamma_{\textrm{t}}(\thetag)\equiv-\Re\left(\gamma(\thetag)\,{\textrm{e}}^{-2{\textrm{i}}\varphi}\right).\end{equation}
 The choice of $U(\theta)$ is arbitrary provided it has a zero mean.
In the following, we will use \cite{2003astro.ph..5089V}

\begin{equation}
U(\theta)\equiv{\frac{9}{\pi\theta_{c}^{2}}}\left(1-\left({\frac{\theta}{\theta_{c}}}\right)^{2}\right)\left({\frac{1}{3}}-\left({\frac{\theta}{\theta_{c}}}\right)^{2}\right).\label{eq:U:filter:def}\end{equation}
 For this particular choice, the variance of the convergence is expressed
in term of the shear power spectrum as 

\begin{equation}
\langle M_{\textrm{ap}}^{2}\rangle={\frac{288}{\pi}}\int{\textrm{d}}\ell\;\ell\; P_{\kappa}(\ell)\left[{\frac{J_{4}(\ell\theta_{c})}{\ell^{2}\theta_{c}^{2}}}\right]^{2}.\label{eq:mapstat:def}\end{equation}
 The variance of the aperture mass is therefore a broad-band estimate
of the convergence power spectrum given in Eq.(\ref{eq:gfunc}), which
can directly be estimated from the galaxy shapes. This is the quantity
which will be used to study the evolution of dark energy in section
\ref{sec:results}. In order to make predictions on cosmic shear observable,
one only needs to compute the convergence power spectrum in dark energy.
As shown in Eq. (\ref{eq:pofkappa}), it only depends on the cosmology
through the relation between angular distances and redshifts, and
through the power spectrum of the matter density fluctuations. The
source distribution $p(z)$ can be determined from the data and does
not depend on the cosmology. 

A complete discussion on the computations of the weak lensing power
spectrum with a cosmology with a non-trivial dark energy has been
done in BB01. We only reproduce here the conclusions of this work.

\subsection{Cosmological distances }

The relation between the cosmological distances and redshift is given
by eq. (\ref{eq:cosmo:dist}). The dark energy component only leaves
an imprint on $\chi(z)$ by modifying the acceleration of the universe
obtained by solving Eq. (\ref{eq:fried1}-\ref{eq:fried2}). 

The modification of the relation distance / redshift affects mildly
the convergence power spectrum. It can be summarized into two simple
effects : a normalization change and a scale shift (similar to the
modification of the position of peaks in CMB). 

The lensing efficiency function (Eq. \ref{eq:gfunc}-\ref{eq:gfunc:pz})
acts as a selection window. It is maximum for lenses located roughly
at mid-distance between the observer and the source galaxies. This
selection effect can be approximated by replacing the $g(\chi)^{2}$
term in in Eq. (\ref{eq:pofkappa}) by a Dirac function \begin{equation}
g^{2}(\chi)\sim g^{2}(\chi_{\textrm{mid}})\delta(\chi-\chi_{\textrm{mid}}).\label{eq:gapprox}\end{equation}
 In this approximation, the normalization change advertised earlier
is driven by the change in the position of the maximum of the selection
function. 

The scale shift is also easily understood with this approximation.
This shift comes from the $P_{3D}(\frac{\ell}{\mathcal{D}(\chi)})$
term in Eq. (\ref{eq:pofkappa}). The modification of the maximum
of the efficiency window selects a different depth for projection. 

In the following, we will show that the matter power spectrum can
be split into two evolution regimes. At large scale, the linear regime
is well described by a power law. The effect of dark energy can be
completely re-absorbed into a change in normalization. At small scales
however, in the non-linear regime, the power law approximation is
no longer a good description of the power spectrum. Like the modification
of the peaks in the CMB power spectrum, the scale under which is seen
the transition from linear to non-linear regime will be slightly shifted
by the above effect.

\subsection{Power spectrum of matter\label{sub:Power:spec}}

From a practical point of view, there is no need to compute the whole
power spectrum of the density fluctuation. Only a narrow range of
scale, from a few arc-second (galaxy scale) to a few hundreds of arc-minutes
across the sky is enough. Large scales ($>~5$ degrees) are difficult
to access observationally anyway; the weak lensing correlation amplitude
is small, where the residual systematics might be a problem, and the
surveys capable of such measurement are not yet planned. At scales
smaller than a few arc-seconds, the number of lensed galaxies drops,
and the noise blows up. 

At redshift one, a few degrees corresponds to a few hundreds of Mpc,
which is far below the horizon size at recombination. As stated in
section \ref{sub:Cosmology} we can safely assume that for the scales
of interest, a classic CDM power spectrum is a good approximation
of the power spectrum of the matter density fluctuation. Moreover,
for these scales, the power spectrum behaves essentially as a power
law. The remarks made above on the cosmological distances holds, and
the effect of the modification of the equation of state of dark energy
on the cosmological distances translate, for this range of scales,
in a simple normalization shift. 

We have yet to investigate the evolution of the power spectrum from
recombination until now. The growth of structures is modified by the
presence of a dynamic dark energy component. At the linear order,
it is given by the well known equation\cite{1993ppc..book.....P}\begin{equation}
\ddot{D}_{+}(t)+2H\dot{D}_{+}(t)-\frac{3}{2}H^{2}\,\OmM(t)\, D_{+}(t)=0.\label{eq:growthlinear}\end{equation}
 In this equation, the matter acts as a source term that increases
the depth of the potential well and tends to increase the density
contrast. On the opposite, the expansion of the universe acts, via
the second term, as a friction effect and reduces the efficiency of
gravitation to increase the density contrast. This term carries all
the effect of dark energy on the growth of structures. 

At high redshift, the dark energy is completely dominated by the radiation
or the matter density. During the radiation era, the growing mode
of equation Eq. (\ref{eq:growthlinear}) can be obtained analytically\begin{equation}
D_{+}\propto a^{\frac{3}{2}}.\end{equation}
 During matter domination, the solution is also well known\begin{equation}
D_{+}\propto a.\end{equation}
 These remarks allow us to integrate Eq. (\ref{eq:growthlinear})
in our model; we do not know any other initial condition that would
allow us to perform the integration and be sure to keep the growing
solution. 

It is worth noting here that there is no other way to compute the
growth of structure\cite{BB01,2003astro.ph..5286L}. The well known
integral solution of Eq. (\ref{eq:growthlinear}) is valid when the
universe only contains matter, radiation curvature and a cosmological
constant. It is also easy to convince oneself that there is no solution
to Eq. (\ref{eq:growthlinear}) that can be integrated from today
toward the past. Equation (\ref{eq:growthlinear}) generically admits
a growing and a non-growing solution. Only the first one is of cosmological
interest, and there is no way to build this solution when integrating
from the final point of the evolution. 

\begin{figure}
\includegraphics[%
  bb=90bp 70bp 770bp 560bp,
  clip,
  width=1.0\columnwidth]{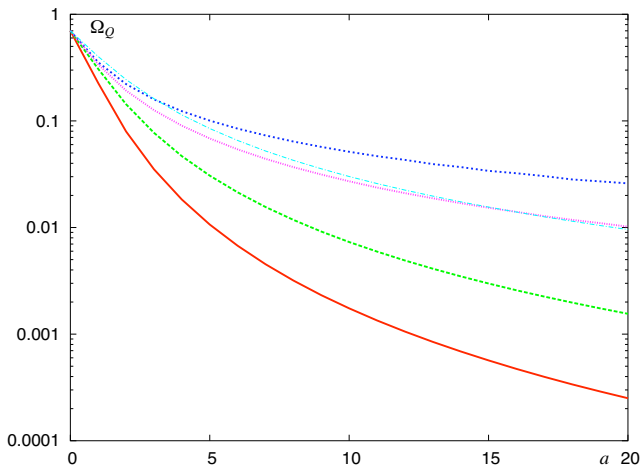}

\caption{\label{cap:Q:z}The energy density of dark energy normalized to the
critical density as a function of redshift. Thick plain line is the
classic $\Lambda$ model. The thick long dashed line and thin dot
dashed line are resp. a $w_{Q}=-0.8$ and $w_{Q}=-0.6$ models. The
short dashed and dotted line are resp. $w_{0}=-0.8,\: w_{1}=0.2$
and $w_{0}=-0.8,\: w_{1}=0.3$ models (complete description of the
parameterization can be found sec. \ref{sub:Dark:Model}). The sooner
the dark energy gets close to one, the sooner it will affect the expansion
and the growth of structure. As expected, models with a equation of
state different from $w_{Q}=-1$ contribute significantly to the acceleration
sooner. Models with a varying equation of state contribute yet sooner.
A constant $w_{Q}=-0.6$ model interpolates between the two $w_{1}\neq0$
ones. }
\end{figure}

During the expansion, the growth follows the radiation, and then the
matter solution. When dark energy gets closer to the energy density
of matter, the friction term grows compared to the  source one. The efficiency of gravitational
collapse to build up the density perturbation decreases and the growth
of structures is damped. For the set of models where the dark energy
happens to dominate earlier, this reduction of
the growth rate is experienced at a higher redshift. The exact starting point of
this damping depends on the evolution properties of the dark
energy model. Models with $w_{Q}>-1$ experience this effect earlier
than for $w_{Q}=-1$. For a constant equation of state, the energy
density of the dark energy goes as \begin{equation}
\Omega_{Q}\propto a^{-3(1+w_{Q})}.\end{equation}
 As said above, the $w_{Q}=-1$ model is the cosmological constant
case. When $w_{Q}>-1$, $\Omega_{Q}$ grows as the scale factor $a$
goes to unity. In this case, the dark energy contribution to the expansion
is significant at a higher redshift than when $w_{Q}=-1$. This is
even more important for varying $w_{Q}$, as shown figure \ref{cap:Q:z}.
The modelling of the dark energy used in this figure will be described
later. %

\begin{figure}
\includegraphics[%
  bb=90bp 70bp 770bp 560bp,
  clip,
  width=1.0\columnwidth]{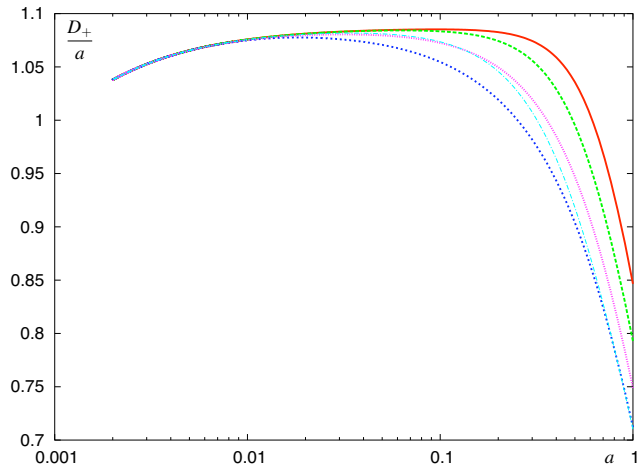}

\caption{\label{cap:dplus}linear growth for different models. Models are
the same than fig. \ref{cap:Q:z}.The growth are normalized to the
recombination era. The modification of the equation of state induce
a precocious acceleration that decrease the efficiency of gravitational
collapse at higher redshift than in the $w_{Q}=-1$ case. A variation
in the equation of state ($w_{1}\neq0$) amplify this effect. As expected
from fig. \ref{cap:Q:z} a constant equation of state model can partly
mimic a varying equation of state: if one knows the CMB normalization
and today normalization of the fluctuation of structures, one cannot
distinguish between a $w_{Q}=-0.6$ and a $w_{0}=-0.8,\: w_{1}=0.3$
models.}
\end{figure}

Figure \ref{cap:dplus} shows the result of a numerical integration
of eq. (\ref{eq:growthlinear}) for different models. The damping
of the growth appears earlier in the varying equation of state models,
compared to the constant equation of state. %

For a fixed redshift, the modification of the growth of structure
at linear order will be degenerated with the normalization of the
power spectrum. Of course, following the linear power spectrum along
the line of sight allows to break this degeneracy. Unfortunately,
the measured shear power spectrum is only a projection of the mass
power spectrum along the line of sight. We showed in previous section
that this projection acts like a selection effect on a single redshift
plane. Therefore, the integrated growth effect will be, in the linear
part of the lensing power spectrum, indistinguishable from a normalization
shift. 

At small scales, the evolution of the density contrast changes dramatically.
Virialized objects are formed and evolve in a different regime than
the simple one described by the linear approximation. BB01 showed
that this regime can potentially solve part of the degeneracy described
above. The weak lensing measurement on galaxies allows access to the
bigger scales of this non-linear regime. 

The perturbation approach cannot describe this regime as the density
contrast is very big at the scale of virialized objects. A complete
computation of this regime cannot be done analytically. One has to
rely on hypotheses on the properties of this regime to be able to
describe it. Several {}``classic'' description of this regime has
been proposed (among others see \cite{1991ApJ...374L...1H,1996MNRAS.280L..19P,2002PhR...372....1C}).
We will follow here the choices made in BB01. We assume that the Stable
Clustering Ansatz provide a valid description of the smaller scales
of the non-linear regime. It states that virialized objects are stable,
that is to say that their physical size does not vary with the expansion
of the universe. Hence, at the scale of these objects, the density
contrast has to grow to match exactly the expansion. Instead of a
growth of order $a$ or smaller, the density contrast evolves as $a^{3/2}$. 

One should note that the scales described by the Ansatz are below
the shear measurement scales. The transition between the linear and
non-linear regimes is described by a mapping between the
two regimes \cite{1991ApJ...374L...1H}. This mapping is calibrated
using a fit to n-body simulations, as described in Peacock \& Dodds
\cite{1996MNRAS.280L..19P}. At large scale, the mapping keeps unchanged
the linear regime, and at small scales, it must go as $(a^{2}/g^{2}P)^{3/2}$.
Although this type of mapping has been widely tested for many different
cosmologies, it has never been tested for dark energy models. However,
given the robustness of the mapping for very different cosmologies,
we assumed it remains valid for the class of models studied here. 

This is a very strong assumption. It can partly be justified by the
fact that it is unlikely that a smoothed dark energy component with
no coupling can affect the small scale behavior of the matter. Its
influence should only appear as a change in the expansion and thus,
as we have shown above, as a modification of the linear growth of
structures. The strength of this argument argument is enhanced by
a recent result from another description of the non-linear regime
usually called the \emph{halo model}. This approach describes the
virialized object as dark matter halos of known%
\footnote{read \emph{fitted on N-body simulations}%
} profile and abundance depending on the cosmological parameters \cite{2002PhR...372....1C}.
The results and concepts behind this approach have been successfully
tested in the context of dark energy \cite{2003astro.ph..3304K,2003astro.ph..5286L}.
In particular the differences observed between halos in $\Lambda$
cosmologies and in cosmology with non trivial dark energy can be explained
by an earlier entrance into the non-linear regime. The observed discrepancies
are, as expected, all explained by the modification of the linear
growth of structures \cite{BB01,2002AA...396...21B}. 

As we just emphasized above, the effect of a non trivial dark energy
on the non-linear regime are all encoded in the linear growth. As
the structures grow slowly, they reach the non-linear regime earlier
(for the same final normalization). Thus, changing their growth at
a higher redshift they undergo a longer non-linear evolution. This
translates, in the context of Stable Clustering Ansatz into a higher
asymptote of the power spectrum at large wave numbers \cite{BB01},
and in the halo approach, into a halo profile with more substructure
and more mass in the center \cite{2003astro.ph..3304K}. Note that
the case described here is very similar to a comparison between an
open and a flat model. 

This will show up in the weak lensing power spectrum in two ways.
First, the earlier entrance in the non-linear regime, once projected,
gives a transition between the linear and non linear regime at smaller
$\ell$. Second, the amplitude of the power spectrum at small scale
is expected to be higher. BB01 proposed estimations of these two effects.
In particular, due to the different evolution in the non-linear regime,
the modification of the asymptote height is expected to go as the
third power of the normalization modification in the linear regime. 

Probing the small scales of the shear power spectrum will reduce the
degeneracy between the determination of dark energy properties and
the normalization of the matter density fluctuation. The extend to
which this can be done is the main subject of sec. \ref{sec:results}. 

Before going to next section, we would like to emphasize the fact
that any modification in the nature of dark matter will be degenerated
(at the level of the growth of structures of course) with the effect
of dark energy. In particular, inclusion of hot dark matter will also
modify the rate of growth. It is expected that this modification should
decrease the amount of small scale structure, thus suppressing the
effect of dark energy. We are then likely to underestimate the effect
of dark energy in those models.

\subsection{Dark Energy Model \label{sub:Dark:Model}}

The evolution of $w_{Q}$ is \emph{a priori} free. It has to be fixed
by a proper model of dark energy. Several models have been studied.
The simplest being the \emph{minimal quintessence model}, where $w_{Q}$
is constant. Another very interesting class of models are the so-called
\emph{tracking potential models}. 

These models have been extensively described \cite{1999PhRvD..59l3504S}.
Their interest lie in the fact that their equation of state parameter
is constant during most of the universe evolution. The constant equation
of state is an attractor solution for $w_{Q}$ when the expansion
of the universe is dominated by another component (like the radiation
or the matter). Of course, when the dark energy reaches the order
of magnitude of the other energy densities, it leaves its attractor
evolution. This attractor ensures that the initial conditions of dark
energy does not have to be finely tuned. Whatever is the starting
value of the dark energy%
\footnote{Usually the initial condition are free within a few tens of order
of magnitude. %
}, it has to reach the attractor and will always exit the domination
of matter at the same point, which in turn depends of the exact model. 

This explains the interest these models have met among the high energy
physics community. In particular, it has been shown that some tracking
potential models can be built within particle physics models. For
example, P. Brax and J. Martin \cite{2000PhRvD..61j3502B} proposed
a version of the Ratra-Peebles model \cite{1988ApJ...325L..17P} that
can be embedded in super-gravity models. 

Minimal and tracking models are not the only dark energy models available.
The problem however is that it does not exist any common framework
to describe the different dark energy models that allows a direct
comparison between them as we plan to do here, in the context of weak
lensing. We attempt here to provide a suitable parameterization that
will allow to constrain the dark energy properties using shear measurements.
This is a simplification of the problem: we only have to model the
evolution of dark energy that can leave an imprint on the weak lensing
power spectrum. As a consequence, we only have to consider its impact
on the growth of structure and on the relation distance-redshift. 

We choose to parameterize the evolution of dark energy in term of
its equation of state. This choice is the most prevalent one. This
is by no mean the only possible parameterization \cite{2003astro.ph..3009A}.
As stated in section \ref{sub:Cosmology}, the knowledge of the EOS
of dark energy is sufficient to solve Eq. (\ref{eq:fried1}-\ref{eq:fried2})
and to compute $\chi(z)$ and $D_{+}$. 

If we do not make any further assumptions, $w_{Q}$ can freely vary
between $-1$ and $1$. The easiest solution is to assume that $w_{Q}$
can be written as a power series of the redshift \begin{equation}
w_{Q}=\sum w_{i}\, z^{i}.\end{equation}
 This is the way most SNIa data are analyzed. More precisely, using
the fact that the data collected relates directly to the integral
of the expansion factor up to a redshift of order one, it is enough
to only investigate this equation of state in a perturbative development\begin{equation}
w_{Q}=w_{0}+w_{1}z\dots\label{eq:snia:eq:st}\end{equation}
 The possible determination of the two first orders of this development
\cite{2001PhLB..500....8A,2001AA...380....6G,2001PhRvL..86....6M}
as been studied. 

This approach is not valid in our case. As noted above, one can only
compute the growth of structure from recombination. A perturbative
development as Eq. (\ref{eq:snia:eq:st}) is of course not suitable
for our purposes; it leads to arbitrarily growing equation of state,
which is not physical! The full power series is also useless. As explained
above, it is in the transition between linear and non-linear scales
that one can expect to gain some information on the dark energy. As
shown figure \ref{cap:Q:z} and \ref{cap:dplus}, one can expect some
level of degeneracy even between a given varying and a non varying
EOS models. Accordingly, it is very important here that we greatly
reduce the parameter space allowed to $w_{Q}$. 

Since we obviously cannot explore completely the equation of state
space if we want to restrict ourselves to a small number of parameters,
we have to make some assumptions on the behavior of the dark energy
equation of state. There are already too many different models of
dark energy, and it is impossible to describe them with just one simple
parameterization. Attempts have been made to generically describe
dark energy with a simpler parameterization than the naive power series.
They however produce results with yet to many parameters \cite{2002astro.ph..5544C}
for our purpose. Another possible approach would be a principal component
analysis designed for shear power spectrum inspired from ideas proposed
in \cite{2002astro.ph..7517H}. 

Here, to reduce the complexity, we add a physical assumption. We will
only be interested in models who exhibit a behavior similar to the
one of the tracking potential models. This is a very strong assumption.
If the choice of model is hard to justify, we will however give here
a few arguments in favor of the tracking potential behavior. 

The behavior of the dark energy equation of state at large redshift
is in fact quite irrelevant for us. Indeed, tracking models assures
that the equation of state of the dark energy is constant as long
as it is dominated by the other components \cite{1999PhRvD..59l3504S}.
When dark energy is dominated its variation have little or no impact
on the expansion of the universe. The growth of structure is not greatly
modified by variation of $w_{Q}$ when the dark energy component does
not contribute significantly to the expansion. Accordingly, assuming
a constant or varying equation of state during domination of matter
or radiation is not relevant to our problem. Of course if one assumes
that dark energy can be dominant at high redshift, this discussion
is not valid. However, an such a dark energy model would leave a huge
imprint on the CMB and would be most probably already ruled out by
observations. %
\begin{figure}
\begin{tabular}{c}
\includegraphics[%
  bb=80bp 70bp 760bp 560bp,
  clip,
  width=1.0\columnwidth]{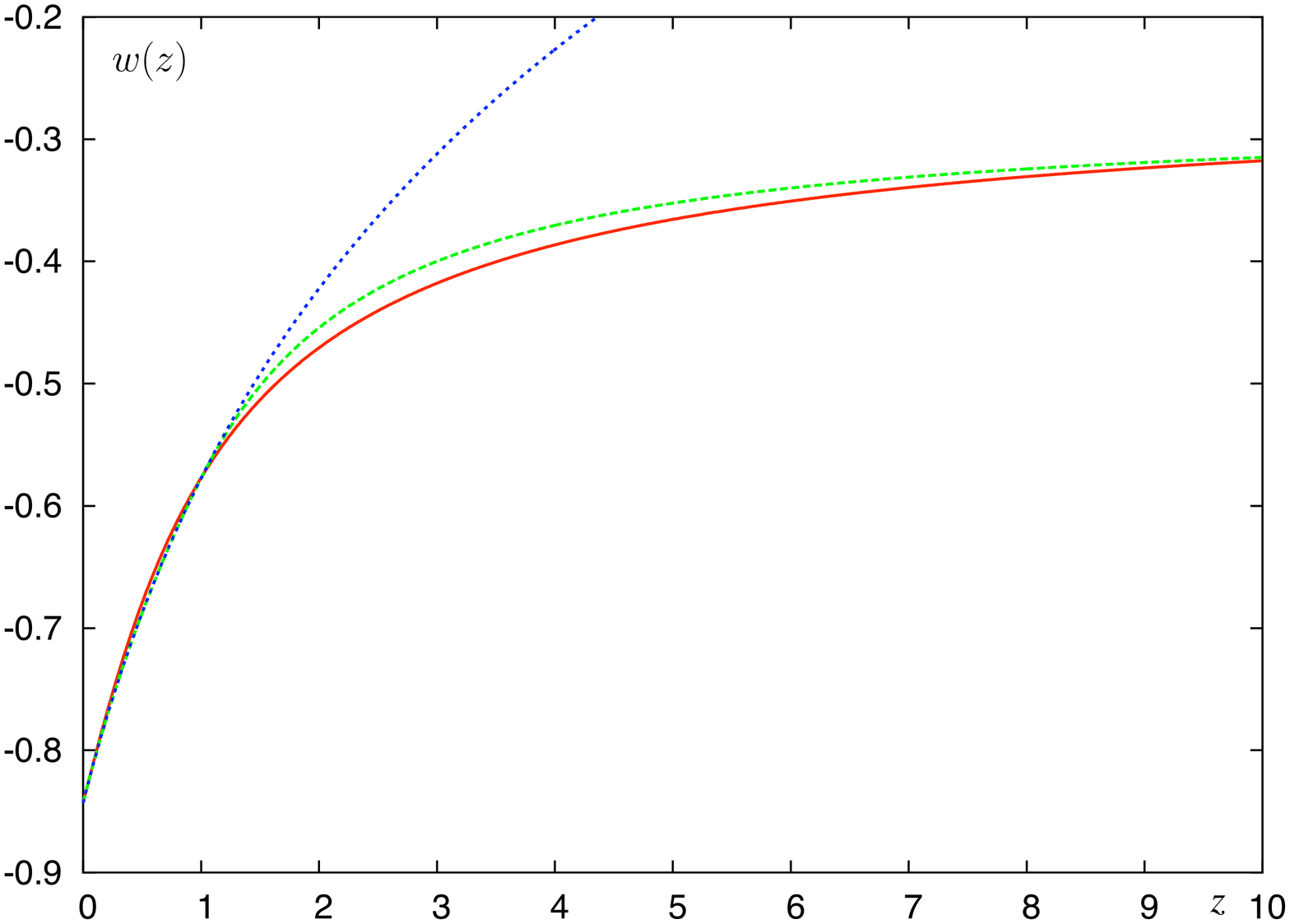}\tabularnewline
\includegraphics[%
  bb=80bp 70bp 760bp 560bp,
  clip,
  width=1.0\columnwidth]{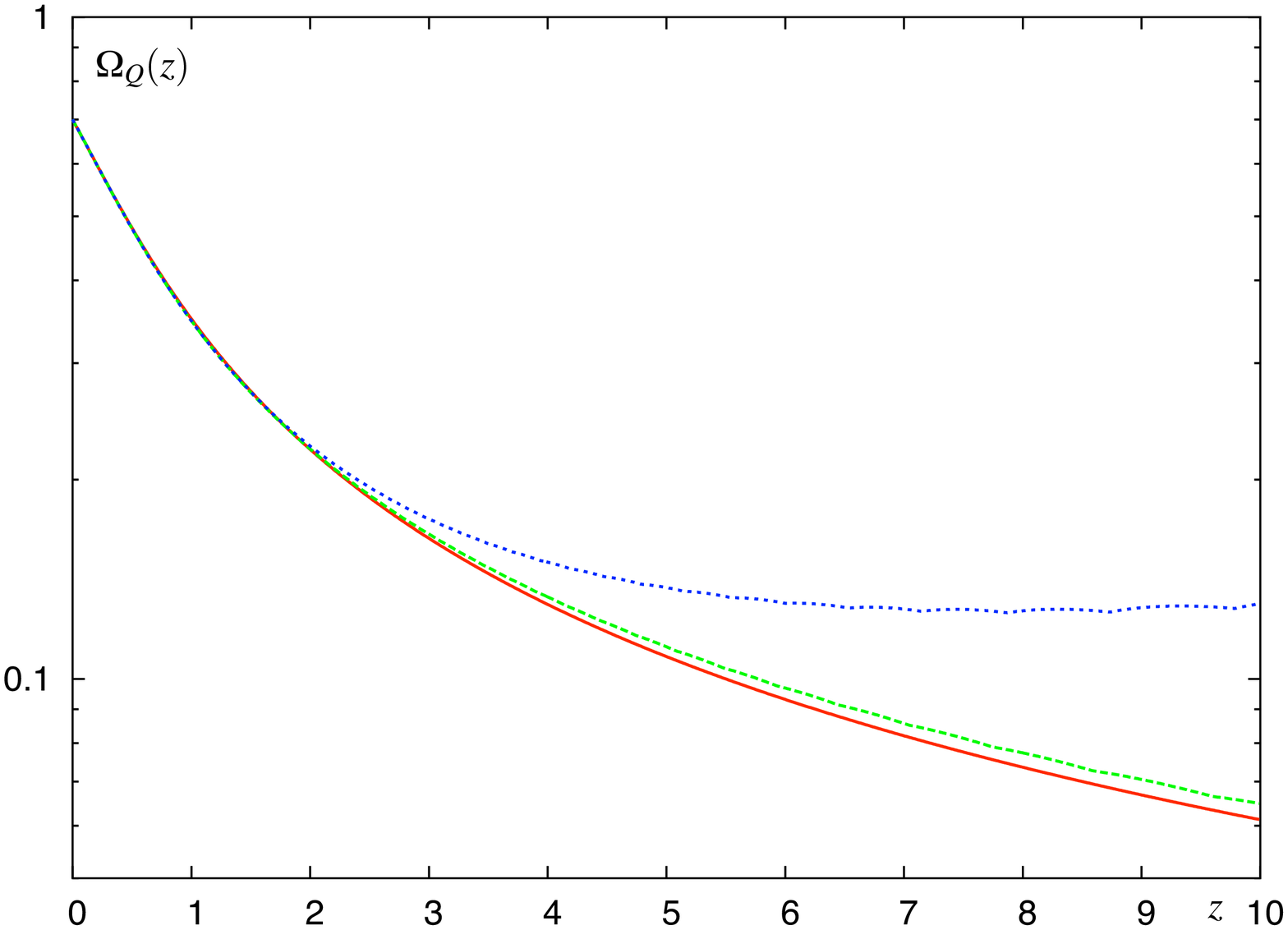}\tabularnewline
\end{tabular}

\caption{Comparison between an explicit SUGRA model and its parameterization.
The equations of state are presented on the top panel, whereas the
energy density, normalized to the critical density are on the bottom.
Plain line is the SUGRA $\alpha=6$ model, long dashed is the logtan
parameterization, short dashed, the log parameterization (see Eq.
(\ref{eq:efst:def}) and (\ref{eq:tan:def})). The parameter $w_{0}=-0.84$,
$w_{1}=0.32$ are measured on the SUGRA model. The log parameterization
quickly fails to fit the equation of state above $z\sim1$. It keeps
a relatively good agreement on the dark energy density the to a higher
redshift. It is not unexpected as the dominant contribution to $H^{2}$
is already the matter energy density. Thus slight variation on the
equation of state of the dark energy are softened on $\Omega_{Q}$.
The discrepancy, however build up very quickly to a factor 2 around
$z\sim8$. While not being in perfect agreement with the SUGRA model,
the logtan parameterization does a better job at following $\Omega_{Q}$.\label{cap:good:job:dimitri}}
\end{figure}

At low redshift, when dark energy reaches the order of magnitude of
the energy density of matter, it will start to contribute to the expansion,
and induce a new period of acceleration. Variations of $w_{Q}$ then
leave a potentially strong imprint on the shear power spectrum, through
modifications of the cosmological distances and structure growth.
This is where our assumption on the shape of the model is important.
Other models of the dark energy can have very different behavior.
Our choice to only consider tracking like models theoretically reduce
the reach of our conclusion, at least in its details. 

It has been shown \cite{1999MNRAS.310..842E} that the equation of
state of tracking models can be fitted at low redshift by a log function
\begin{equation}
w_{Q}\sim w_{0}+w_{1}\log(1+z).\label{eq:efst:def}\end{equation}
 This behavior is roughly valid up to redshift $z\sim1$ at least
for SUGRA and Ratra-Peebles potentials. Note that this equation of
state parameter admits Eq. (\ref{eq:snia:eq:st}) as its Taylor expansion
at small $z$. 

The parameterization given in Eq.(\ref{eq:efst:def}) fails quickly
above $z\sim1$ (see figure \ref{cap:good:job:dimitri} ). It has
to be modified to account for the evolution of the equation of state
for $z>1$. We propose to use an arctangent which has the property
to quickly reach a constant value. We will then use\begin{eqnarray}
w_{Q} & =\label{eq:tan:def}\\
 &  & \hspace{-1cm}\left\{ \begin{array}{l}
w_{0}+w_{1}\log(1+z),\:\textrm{if }z\leq1\\
w_{0}+w_{1}\left[\log(2)-\arctan(1)+\arctan(z)\right],\:\textrm{if }z>1\end{array}\right..\nonumber \end{eqnarray}
 This is a crude assumption, because the value of $w$ at high $z$
will be $w_{0}+w_{1}(\log(2)-\arctan(1)+\pi/2)$ which has no reason
to correspond to the high $z$ asymptote of a given tracking potential
model. However, as said above, the dark energy component is completely
dominated at high redshift by the energy density of matter, and the
expansion is nearly insensitive to its evolution. In the end, the
only thing that matters is the evolution of dark energy from the epoch
when it starts to dominate. This epoch can be at redshift as high
as 10. For example, in a SUGRA $\alpha=6$ model\cite{2000PhRvD..61j3502B}
the energy density of dark energy represents 10\% of the total energy
density as early as redshift $z\sim5$ (see figs. \ref{cap:good:job:dimitri}
and \ref{cap:Q:z} ). 

\begin{figure}
\begin{tabular}{c}
\includegraphics[%
  bb=70bp 60bp 770bp 570bp,
  clip,
  width=1.0\columnwidth]{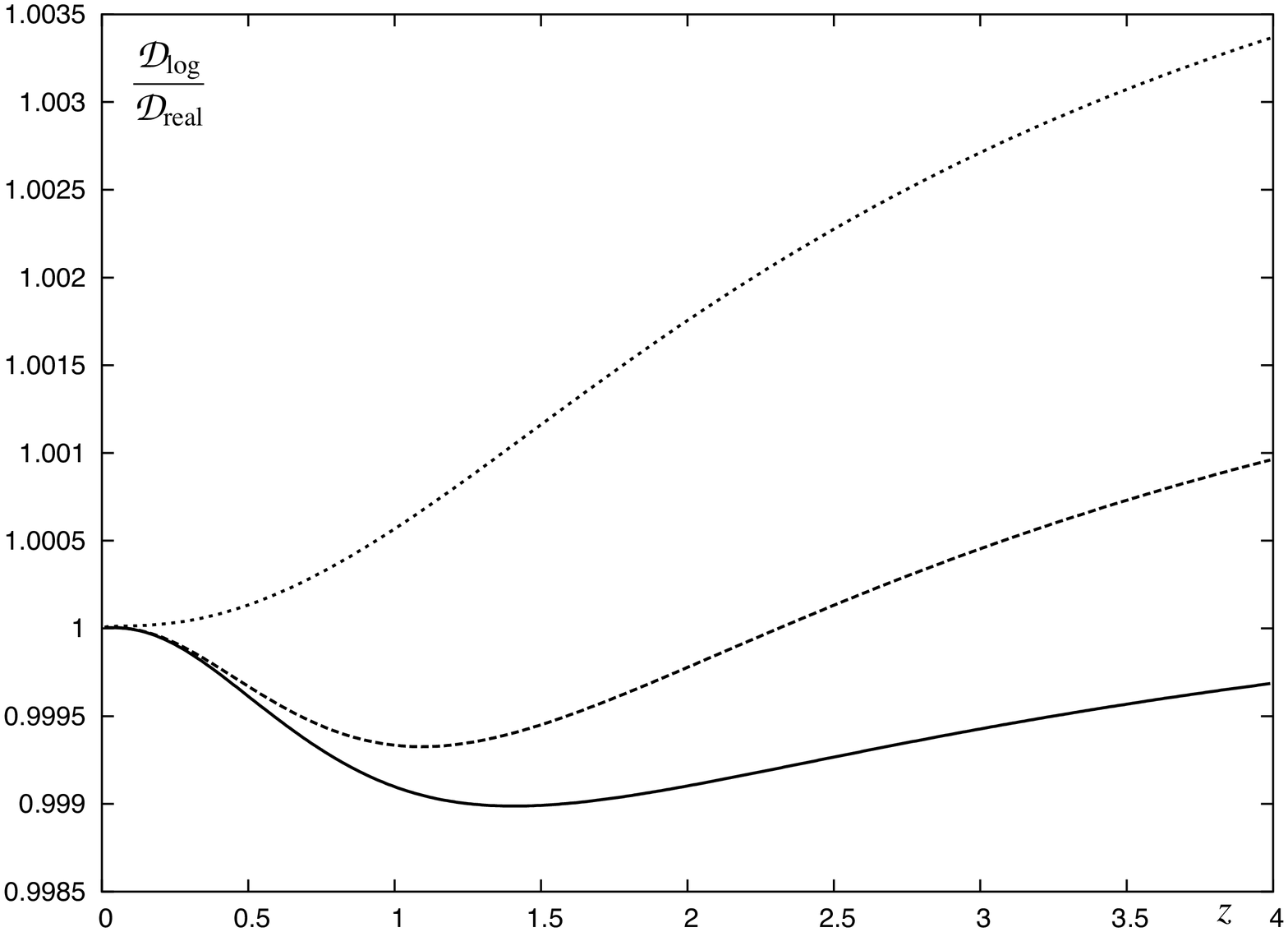}\tabularnewline
\includegraphics[%
  bb=70bp 60bp 770bp 570bp,
  clip,
  width=1.0\columnwidth]{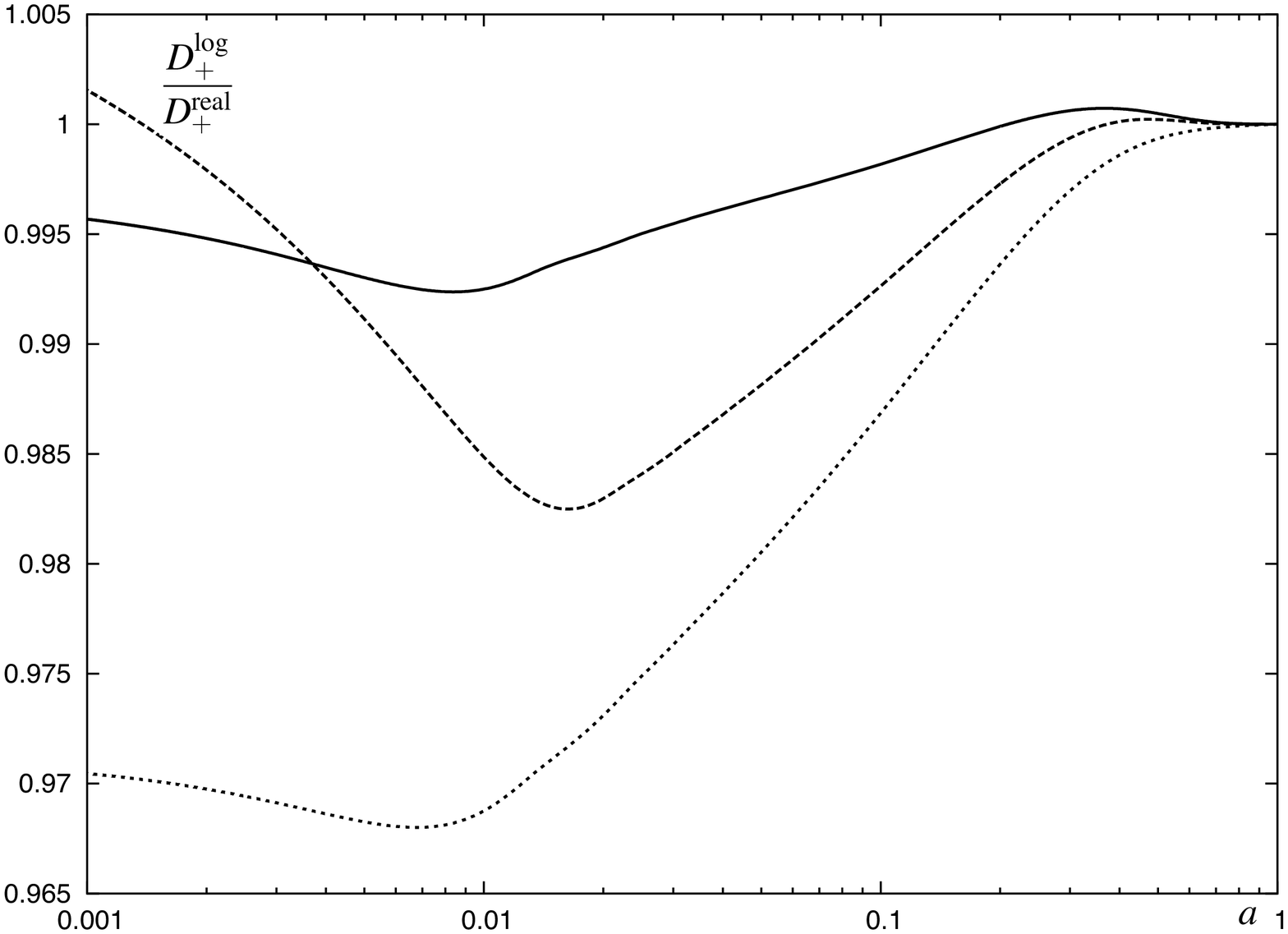}\tabularnewline
\end{tabular}

\caption{\label{cap:distances:dplus}Comparison between real tracking models
and their approximated version using Eq. (\ref{eq:tan:def}). Top
panel present the comoving distances, bottom the growth of structure.
The lines plain line is a SUGRA $\alpha=6$ model, the dashed one,
SUGRA $\alpha=11$ and the dotted a Ratra-Peebles, $\alpha=4$ model.
The discrepancy on the angular distances computed with the real model
and our parameterization is below the percent up to $z=4$. The discrepancy
for the linear growth if of order 3 \% up to the recombination. Our
approximated formula with its very small number of parameters gives
a good approximation of the quantities on which is computed the weak
lensing effect. }
\end{figure}

The parameterization, Eq. (\ref{eq:tan:def}), is not very good at
fitting the equation of state. As shown figure \ref{cap:good:job:dimitri}
for the example of a SUGRA $\alpha=6$ model, it reasonably agrees
with the dark energy density predicted in this model. This is not
the relevant comparison anyway, as we should compare the growth and
distances of the different models. This is done fig. \ref{cap:distances:dplus}. 

The agreement with the explicit tracking models we tested is around
3\%. We will assume that this order of magnitude of the error holds
for any other tracking model. 

We use this parameterization in our analysis. Although it is a \emph{ad
hoc} choice, it conveys most of the feature of a tracking model, and
fits them, at least at the level of the growth of structure and the
cosmological distances. It also has the advantage to be described
by only two parameters, $w_{0}$ and $w_{1}$. We have fixed the change
between the log and tan branches to $z_{c}=1$. Small variations of
$z_{c}$ translate into small modifications of the growth of structure.
For example, taking $z_{c}=1.5$ translates into one percent modification
in $D_{+}/a$, comparable with the error on the modeling. Of course,
since we are only interested to lensing effect to redshift around
1, it leads to negligible modifications in the cosmological distances
and the projection effect %
\footnote{We do have a small dependency on redshift higher than 1 through the
broad distribution of the source $p_{s}(z)$. This effect is small
enough to be neglected here.%
}. Finally, since the parameterization (\ref{eq:tan:def}) admits Eq.
(\ref{eq:snia:eq:st}) as its Taylor expansion, our results are directly
comparable with the well advertised SNIa ones \cite{2001AA...380....6G}.

\section{Results\label{sec:results}}

We perform a maximum likelihood analysis of the aperture mass statistic
for a set of dark energy models. The method is well known and has
been formerly described in \cite{2002AA...393..369V}. Section \ref{sub:Parameter:estimation}
describes the models and surveys that will be investigated. Finally,
we present section \ref{sub:Numerical} the numerical results and
a discussion on the degeneracy between the parameters.

\subsection{Parameter estimation\label{sub:Parameter:estimation}}

We know from previous studies that the gravitational lensing by large
scale structures depends mainly on four parameters: the matter energy
density $\OmM$, the mass power spectrum normalisation $\sigma_{8}$,
its slope, and the redshift of the sources \cite{2002AA...396....1S}.
As described above, we can safely ignore modification to the Cold
Dark Matter transfert function \cite{1986ApJ...304...15B} due to dark energy.
We thus use it and describe the slope of the power spectrum by the
parameter $\Gamma$. 

The lensing effect is also sensitive to other parameters, but to a
lower extend. Therefore when studying the impact of the dark energy
on cosmic shear, one has to incorporate the effect of the main parameters
as well. We highlighted above that in the linear regime at least,
dark energy variation should be degenerated with the normalization
of the power spectrum. By keeping the main parameter we will be able
to test other possible degeneracies. 

We assume that one of the main parameters is known. Indeed, the forthcoming
lensing surveys are supposed to provide an accurate measurement of
the distribution of the sources from photometric redshifts. Moreover,
the redshift dependence is very similar to the $\sigma_{8}$ dependence.
Hence, we will assume $z_{s}$ known. A slight error on its value
can be translated in our result in broader $\sigma_{8}$ constraint. 

Our set of free parameters is chosen as ${\textrm{p}}=(w_{0},w_{1},\OmM,\sigma_{8},\Gamma)$.
We deliberately choose a flat prior. The current CMB results are in
very good agreement with a flat geometry \cite{2003astro.ph..2209S}. 

We compute the likelihood ${\mathcal{L}}({\textrm{p}}|\dg)$, where
the data vector $\dg$ is the aperture mass $\langle M_{\textrm{ap}}^{2}\rangle$
as function of scale: 

\begin{equation}
{\mathcal{L}}={\frac{1}{(2\pi)^{n/2}\left|\Sg\right|^{1/2}}}~\exp{\left[-{\frac{1}{2}}\left({\textrm{\dg-\sg}}\right)^{T}\Sg^{-1}\left({\textrm{\dg-\sg}}\right)\right]},\label{eq:likelihood}\end{equation}
 where $\sg$ is the fiducial model vector and $\Sg:=\langle\left({\textrm{\dg-\sg}}\right)^{T}\left({\textrm{\dg-\sg}}\right)\rangle$
is the covariance matrix. The covariance matrix is computed following
the method described in \cite{2002AA...396....1S} assuming the Gaussian
field approximation. In this work we are interested in two surveys:
one is the ground based Canada France Hawaii Telescope Legacy Survey
\footnote{http://www.cfht.Hawaii.edu/Science/CFHLS/%
}, and the other the spatial Super-Novae Acceleration Probe %
\footnote{http://snap.lbl.gov/%
}. The observational properties of the lensing survey associated with
these two projects are summarized in table \ref{tab:tableobs}. %
\begin{figure}
\includegraphics[%
  width=1.0\columnwidth]{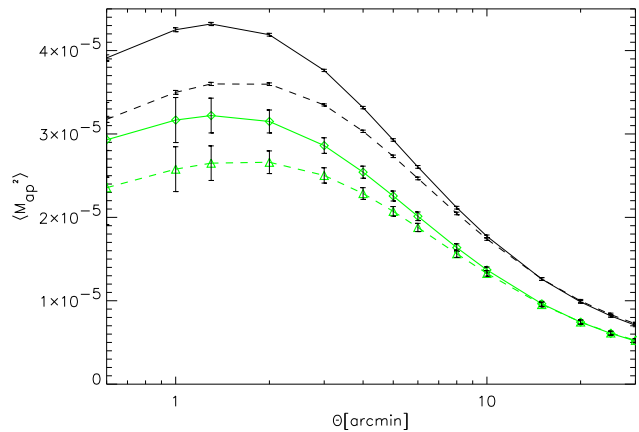}

\caption{Aperture mass variance as function of scale for model3 (solid) and
model2 (dashed), for the CFHTLS (thick bottom lines) and SNAP (thin
top lines). The error bars show the statistical and sampling errors,
assuming aGaussian statistic for the sampling error. \label{fig:models}}
\end{figure}

\begin{table}
{\small }\begin{tabular}{lcccc}
\hline 
&
$\bar{z}_{s}$&
$\theta_{\textrm{deg}}^{2}$&
$n_{\textrm{gal}}$&
$\sigma_{\textrm{e}}$\tabularnewline
\hline
CFHTLS&
0.9&
1790.&
20.&
0.44\tabularnewline
\hline
SNAP&
1.2&
300.&
100.&
0.32 \tabularnewline
\hline
\end{tabular}

\caption{{\small Lensing surveys that will be part of the CFHTLS and SNAP
projects (see text). Entries are source mean redshift $\bar{z}_{s}$,
survey total area $\theta_{\textrm{deg}}^{2}$, source galaxy number
density (per arc-min$^{2}$), and intrinsic ellipticity dispersion
$\sigma_{\textrm{e}}$. \label{tab:tableobs}} }
\end{table}

For the two surveys, we selected three fiducial models (with a cosmological
constant $\Omega_{\Lambda}=0.7$): 

\begin{itemize}
\item model1: ${\textrm{p}}_{1}=(-1,0,0.3,0.9,0.24)$
\item model2: ${\textrm{p}}_{2}=(-0.8,0,0.3,0.9,0.24)$
\item model3: ${\textrm{p}}_{3}=(-0.8,0.32,0.3,0.9,0.24)$
\end{itemize}
The first model is a pure cosmological constant case. Second is a
minimal dark energy model with no variation of the equation of state.
This type of model is widely used in the literature. From the discussion
of section \ref{sub:Power:spec}, it is expected that this kind of
model under-evaluate greatly the effect of a varying EOS with identical
final value. Last model has a varying EOS. The value of $w_{1}$ has
been choose so as to agree with a $\alpha=6$ SUGRA model. It correspond
to a strongly evolving equation of state model. Models with a smaller
$w_{1}$ between our choices interpolate between our model2 and model3. 

We also have to make a choice on the range of parameters we want to
investigate. Maximum likelihood analysis with five parameters is already
a computationally expensive task. It can be reduced in part by narrowing
the range of the parameters and the number of points in each direction. 

For the CFHTLS analysis, it is expected that we will mildly constrain
the parameters. We thus used a relatively sparse grid and relatively
wide parameter ranges. The matter density $\OmM$ will be allowed
to vary between 0.1 and 0.5, while $\sigma_{8}$ will be free between
0.6 and 1.1. The choices for this two parameters are quite conservative.
They allow to probe the full one sigma contour. The slope of the power
spectrum is weakly constrained by the weak lensing measurement we
probe its values between 0.08 and 0.4. The results below (figures
\ref{fig:cfhtls:mod0}-\ref{fig:cfhtls:mod3}) shows that this is
more that enough to correctly probe the parameter space. 

For SNAP, we greatly reduce the range of parameters. The precision
required here forces us to increase the number of computed models,
in particular in the $\OmM,\sigma_{8}$ space. We thus suppose that
it is enough to probe $\OmM$ between 0.28 and $0.32$, and $\sigma_{8}$
between $0.85$ and $0.95$. Nevertheless, it is expected that by
the time SNAP will collect data, previous weak lensing measurements,
CMB, galaxy and cluster surveys will have cut down the accuracy on
this parameters to these levels. We conservatively keep a relatively
wide range on $\Gamma$ ($0.1$ to $0.3$, in agreement with the results
for CFHTLS). 

For both models, we probed the dark energy parameter space between
-1 and 0.6 for $w_{0}$ and 0 and 0.4 for $w_{1}$. Note here that
as described higher, we do not take into account models with an equation
of state more negative than -1. The upper bond on $w_{0}$ correspond
roughly to the degeneracy expected between our target varying equation
of state and a constant equation of state model (see section \ref{sub:Power:spec}).
We do not investigate negative $w_{1}$ models. Negative $w_{1}$
models are very close to the cosmological constant case, and should
be strongly degenerated with it. It is very dubious that weak lensing
will be able to distinguish between them. The $w_{1}$ upper bond
correspond to strongly varying equation of state. It is very difficult
to reach this bond with SUGRA or Ratra-Peebles models.

\subsection{Numerical results -- Discussion \label{sub:Numerical}}

We first compute the aperture mass for our dark energy models. Figure
\ref{fig:models} presents the results for SNAP and the CFHTLS surveys.
It shows that the evolution of the dark energy can lead to a $10$
to $20\%$ effect at small scale. As emphasized section \ref{sub:Power:spec},
this is precisely the expected effect. 

\begin{figure*}
\includegraphics[%
  width=0.80\textwidth]{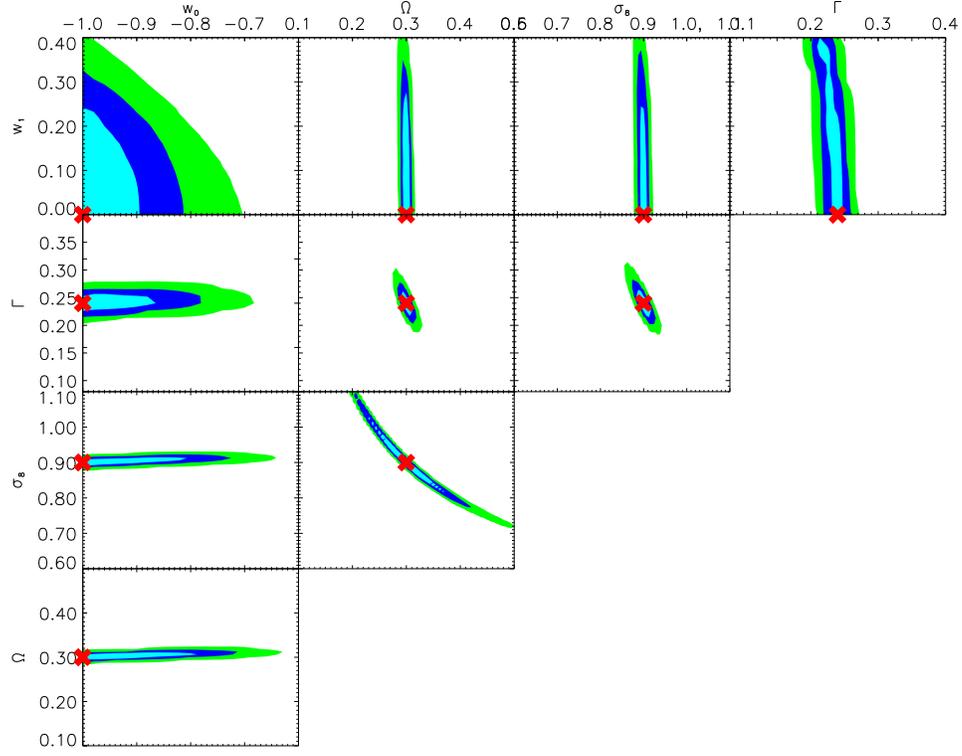}

\caption{Constraints obtained with the CFHTLS survey for a cosmological constant
model (model1). We assumed strong prior for the hidden (not shown)
parameter on each plot. The cross represents the fiducial model. \label{fig:cfhtls:mod0}}
\end{figure*}

\begin{figure*}
\includegraphics[%
  width=0.80\textwidth]{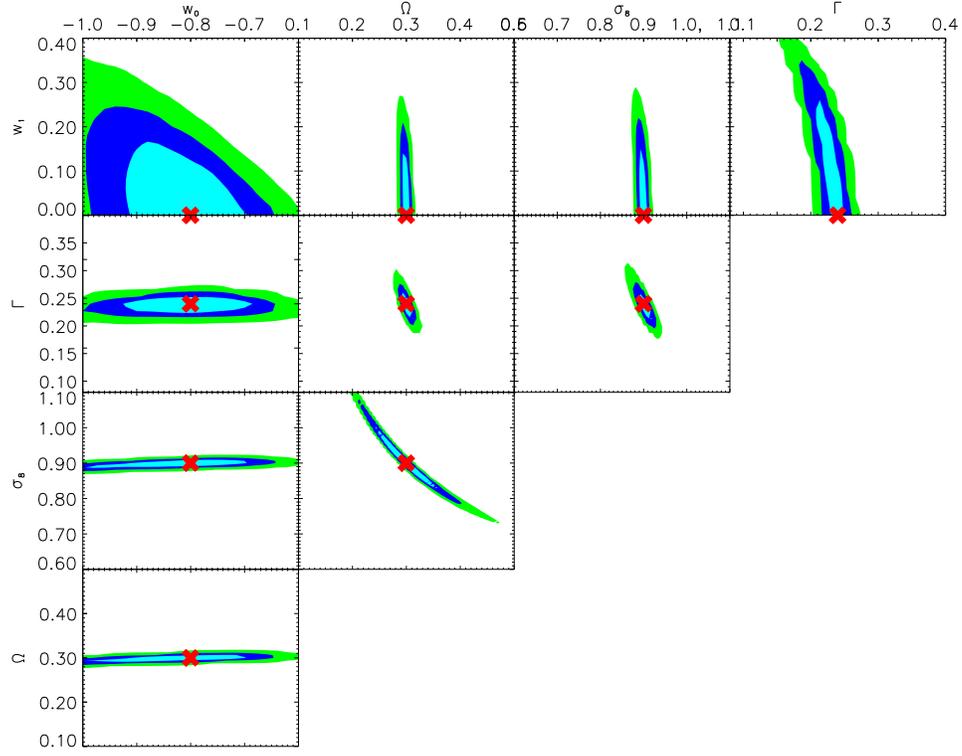}

\caption{Same as Figure \ref{fig:cfhtls:mod0}for a quintessence, non evolving,
model2 (see table \ref{tab:tableobs})\label{fig:cfhtls:mod2}}
\end{figure*}

\begin{figure*}
\includegraphics[%
  width=0.80\textwidth]{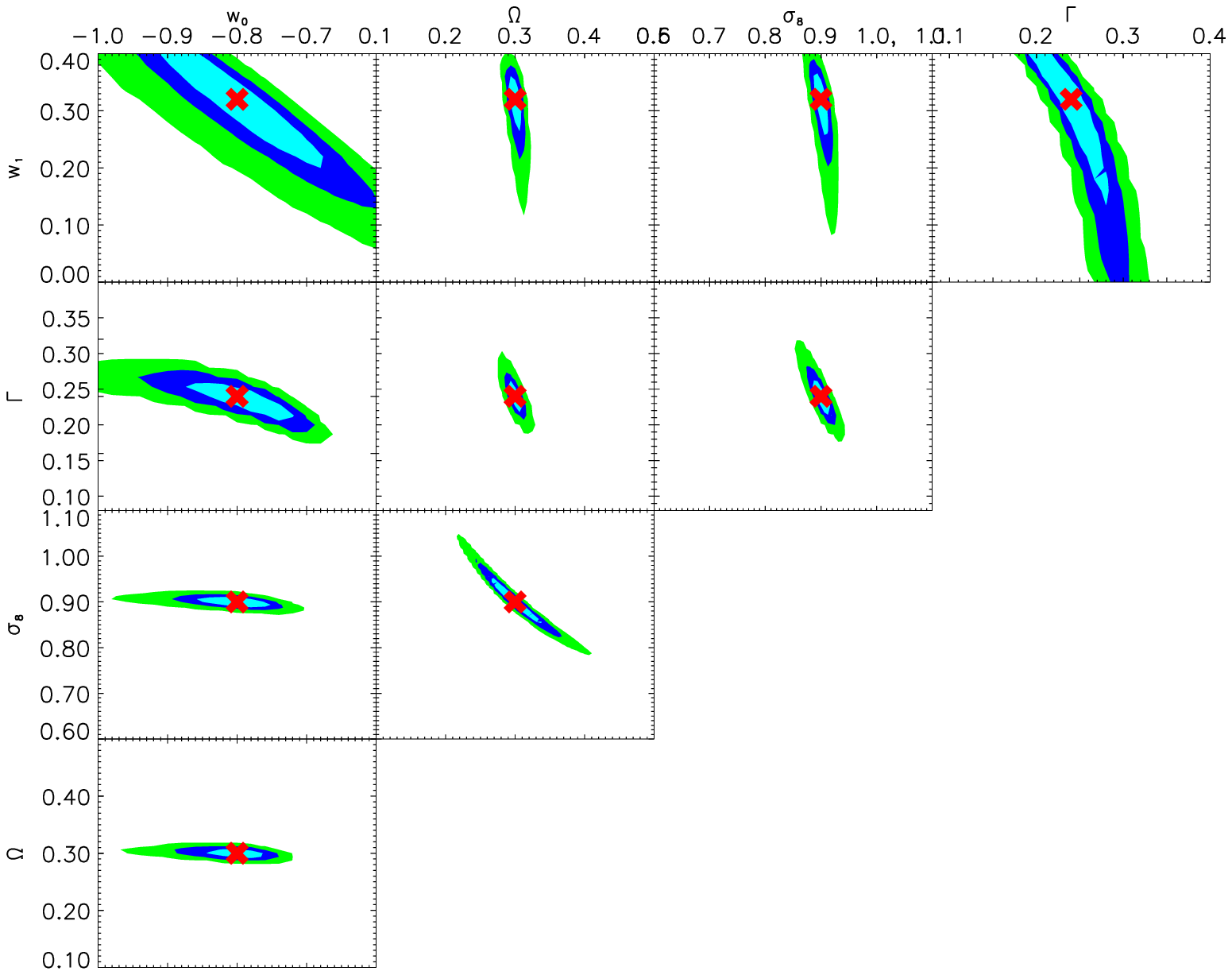}

\caption{Same as Figure \ref{fig:cfhtls:mod2} for an evolving dark energy
model (model3). \label{fig:cfhtls:mod3}}
\end{figure*}

Next we perform the likelihood analysis on our target models. Figures
\ref{fig:cfhtls:mod0}, \ref{fig:cfhtls:mod2}, and \ref{fig:cfhtls:mod3}
show respectively the parameter predictions for the models 1, 2 and
3. All possible combinations of pairs of parameters are plotted in
order to show the direction of degeneracies. On each plot, the two
hidden parameters are assumed to be perfectly known. We first note
the strong degeneracy between the dark energy parameters ($w_{0}$,
$w_{1}$) and the other parameters. The full degeneracy between $w_{1}$
and $\Gamma$ is understood by the fact that the shape parameter describes
the slope of the power spectrum, for a fixed normalisation $\sigma_{8}$.
Changing $\Gamma$ will modify the ratio between linear and non-linear
regime, and the scale of transition. As shown in section \ref{sub:Power:spec},
a change in $w_{1}$ has a similar consequence. 

\begin{figure}
\begin{tabular}{c}
\includegraphics[%
  width=1.0\columnwidth]{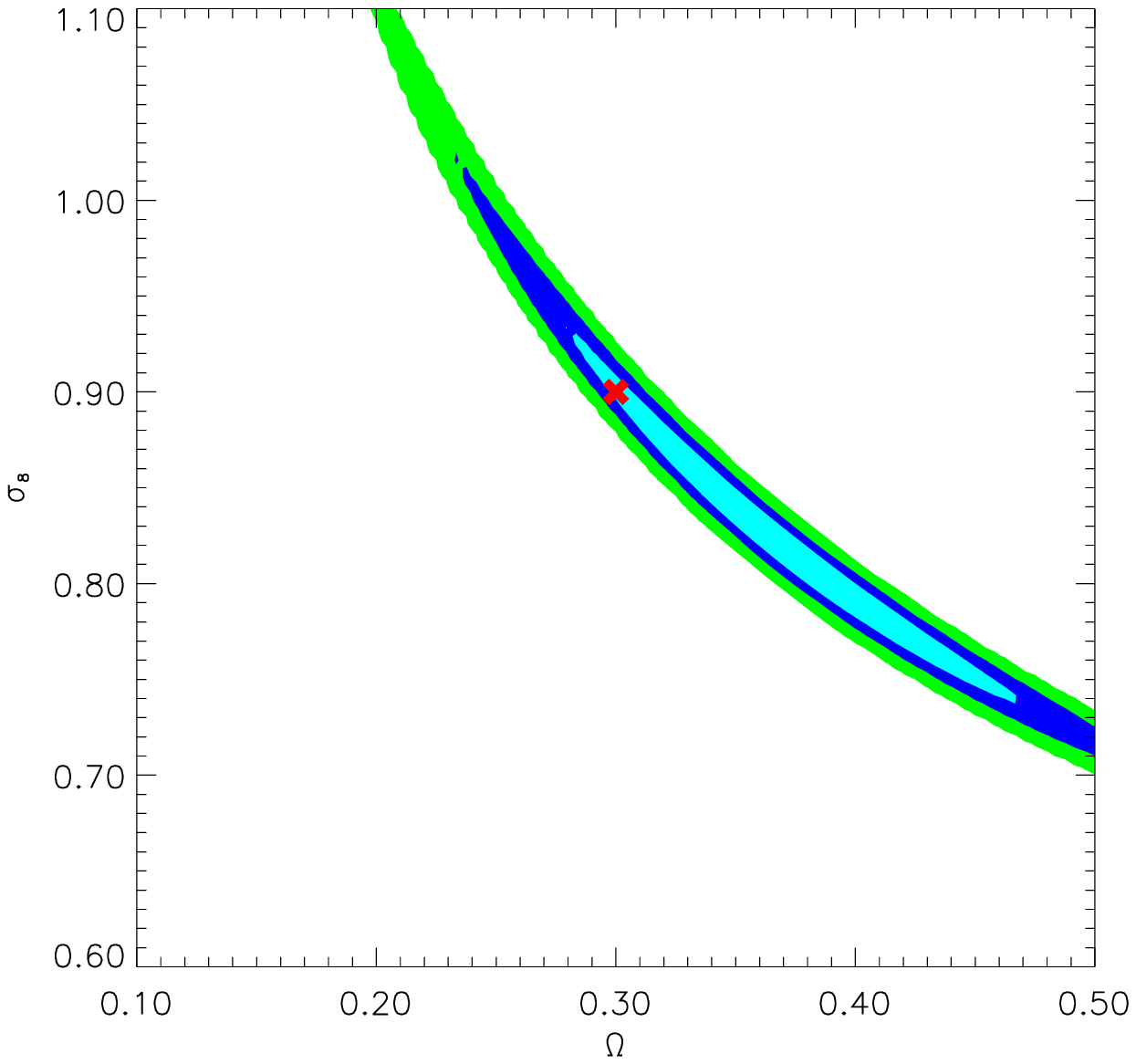}\tabularnewline
\includegraphics[%
  width=1.0\columnwidth]{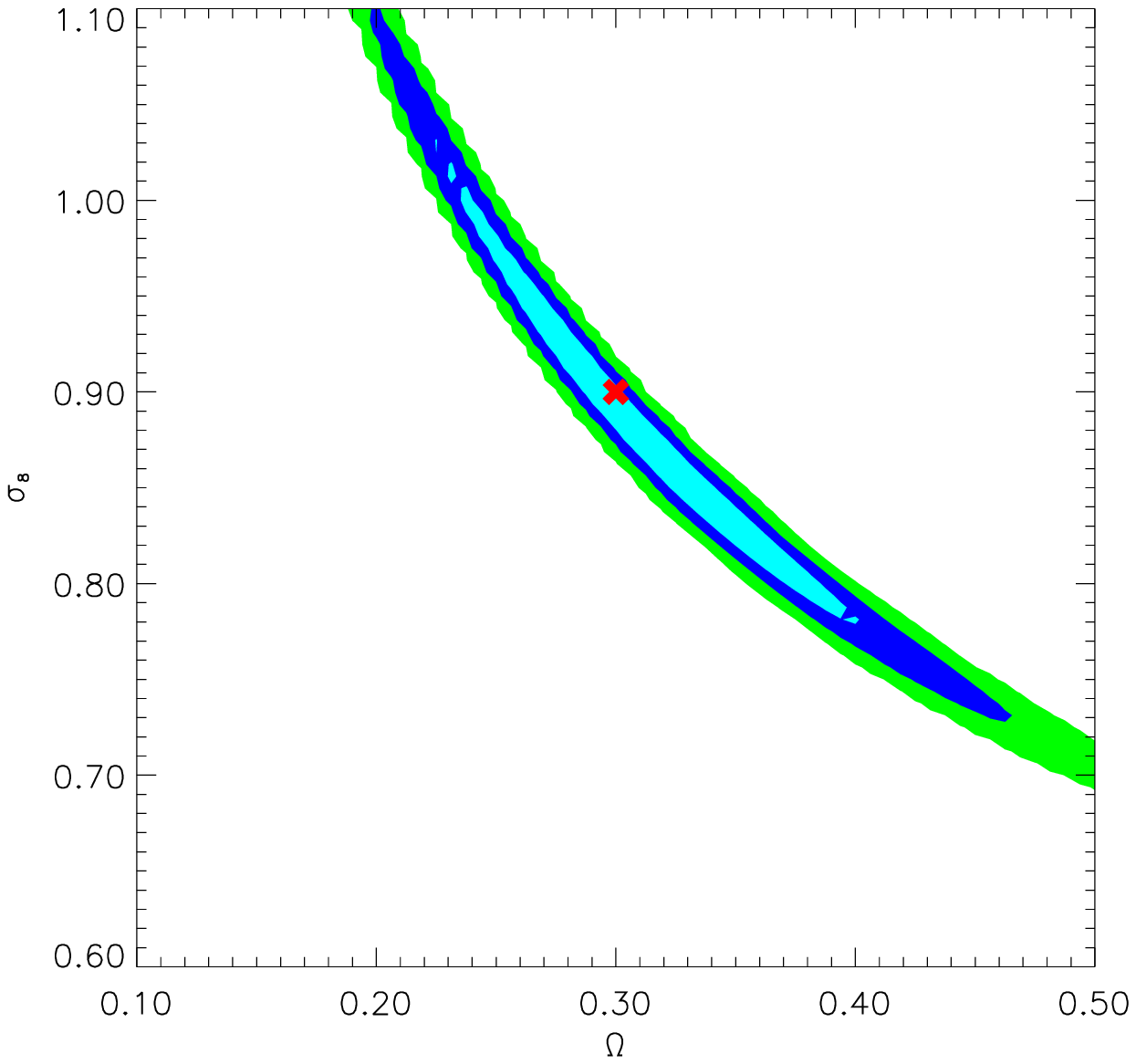}\tabularnewline
\end{tabular}

\caption{Contours in the $\OmM$, $\sigma_{8}$ space when marginalised over
the quintessence $w_{0}\in[-1,-0.7]$ and $w_{1}\in[0,0.4]$. This
is given for the CFHTLS experiment, left panel corresponds to model
1, and right panel to model 2.\label{fig:omega:sigma8}}
\end{figure}

Even allowing for dark energy, the shear two points function remains
a good constraint on $\OmM$ and $\sigma_{8}$. Figure \ref{fig:omega:sigma8}
shows the effect of unknown dark energy parameters (marginalised on
$w_{0}$ and $w_{1}$) on the measurement of $\OmM$ and $\sigma_{8}$.
For the pure cosmological constant model (top panel), we see that
the most probable models correspond to higher $\OmM$ and lower $\sigma_{8}$
than the fiducial model. For the model2 (bottom panel), the normalisation
is underestimated. This figure shows that the width of the $\OmM$,
$\sigma_{8}$ contours is not dramatically affected, but the most
probable models are changed. 

\begin{figure}
\includegraphics[%
  width=1.0\columnwidth]{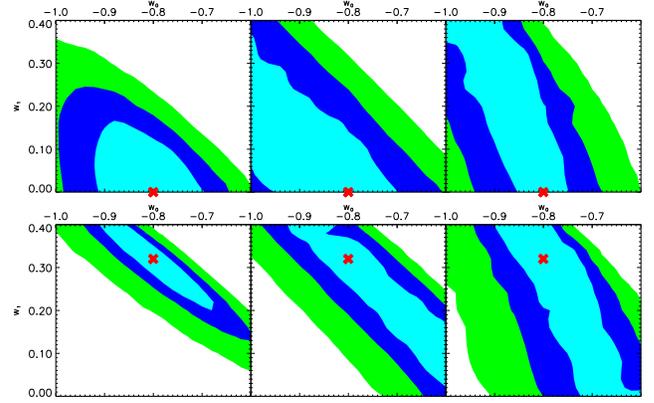}

\caption{CFHTLS constraints with lensing alone on $w_{0}$ and $w_{1}$. Top
panels: model2, bottom panels: model3. Left plot is assuming all other
parameters are known (see Figure \ref{fig:cfhtls:mod2} and \ref{fig:cfhtls:mod3}).
Middle plots is when the mean density and the power normalisation
are marginalised (flat prior) over $\OmM\in[0.1,0.5]$ and $\sigma_{8}\in[0.6,1.1]$.
The right plots show the contour for the marginalisation $\Gamma\in[0.1,0.4]$.
\label{fig:cfhtls:weak:prior}}
\end{figure}

\begin{figure}
\includegraphics[%
  width=1.0\columnwidth]{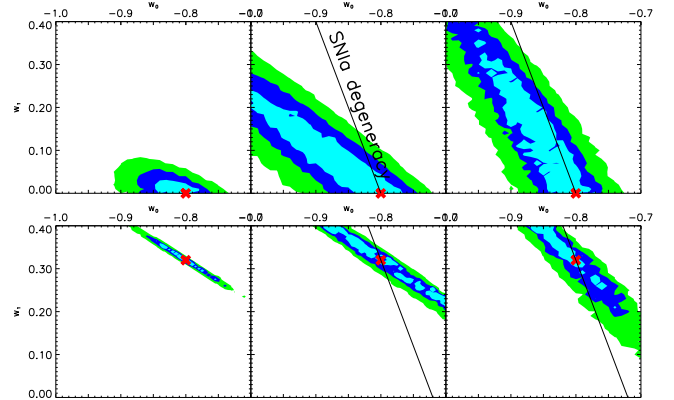}

\caption{Same as Figure \ref{fig:cfhtls:weak:prior}for the SNAP survey. The
marginalisation is performed over the intervals $\OmM\in[0.28,0.32]$
and $\sigma_{8}\in[0.85,0.95]$ for the middle plots and $\Gamma\in[0.1,0.3]$
for the right plots. The line show the direction of degeneracy of
the SNIa (with the supposition of a perfect $\OmM$ knowledge) \label{fig:snap:weak:prior}}
\end{figure}

Super-Novae luminosity surveys have a small sensitivity to the variation
of the equation of state. In particular, it is expected that without
a strong prior on $\OmM$ they cannot provide much information on
$w_{1}$ \cite{2001PhLB..500....8A,2001PhRvL..86....6M,2001AA...380....6G}.
The question is whether the shear two points function also suffers
from this kind of limitation or not. Figure \ref{fig:cfhtls:weak:prior}
and \ref{fig:snap:weak:prior} show the predictions for $w_{0}$ and
$w_{1}$, respectively for the CFHLS and SNAP observations. The left
panels correspond to the contours obtained with a perfect knowledge
of $\OmM$, $\sigma_{8}$ and $\Gamma$. The middle panels are for
a known $\Gamma$, but marginalised over $\OmM$ and $\sigma_{8}$.
The right panels are for known $\OmM$ and $\sigma_{8}$, and marginalised
over $\Gamma$. The top panels are for model 2, and the bottom panel
for model 3. The important result here is that the marginalisation
over $\Omega_{M}$ and $\sigma_{8}$ do not increase too much the
width of the contours, it only restores a degeneracy between $w_{0}$
and $w_{1}$. %

Contrary to the angular diameter distance tests, the weak lensing
is sensitive to the evolution parameter $w_{1}$. The marginalisation
over $\Gamma$ restores the degeneracy along a different direction,
but still does not increase the contours width significantly. It means
that even with a limited knowledge on external important parameters,
it is still possible to constrain the quintessence, in particular
when it evolves with time. In that case indeed (i.e. $w_{1}\ne0$),
the increase of the lensing signal is large enough to allow the CFHTLS
observations to rule out a pure cosmological constant case. However,
an accurate joint measurement of the quintessence parameters and the
others is not possible using the lensing power spectrum alone, because
of the strong degeneracy between $\OmM$ and $\sigma_{8}$. 

This degeneracy is broken with the SNAP lensing survey: according
to Figure \ref{fig:snap:weak:prior}, one sees that cosmic shear observations
alone with the SNAP satellite, provide constraints which are competitive
with the SNIa constraints from the same satellite. The expected constraints
from SNIa alone, assuming a perfect knowledge of $\OmM$, is sketched
on this figure (solid line). It shows that SNIa are less sensitive
to $w_{1}$ than weak lensing. Therefore a combination of SNIa and
cosmic shear could simultaneously probe the dark energy and its evolution.
More precisely, figures \ref{fig:cfhtls:mod0}, \ref{fig:cfhtls:mod2},
and \ref{fig:cfhtls:mod3} show that the knowledge of $w_{0}$ is
irrelevant for constraining $\OmM$ from cosmic shear. On the other
hand, the SNIa measurements are degenerate between $w_{0}$ and $\OmM$.
A combination of the two experiments provide a simultaneous measure
of $w_{0}$ and $\OmM$ without the need for an external measurement
of $\OmM$. We can then use the lensing constraints on $w_{0}$ and
$w_{1}$ (Figures \ref{fig:cfhtls:weak:prior} and \ref{fig:snap:weak:prior})
to estimate the dark energy evolution $w_{1}$. In fact even a poor
knowledge of $\OmM$ can be tolerated; we known from \cite{2001AA...380....6G}
that a marginalisation over $\OmM$ of the SNIa measurements increases
the $w_{0}$, $w_{1}$ contours perpendicularly to the increase of
the same contours from cosmic shear with poor knowledge on $\OmM$
(Figures \ref{fig:cfhtls:weak:prior} and \ref{fig:snap:weak:prior},
middle panels). Adding the cosmic microwave background over-constrains
the parameter space, because the contours in the $\OmM$, $w_{0}$
space are 'perpendicular' to the SNIa and cosmic shear constraints
\cite{2003astro.ph..2209S}. Weak lensing, cosmic microwave background
and SNIa provide therefore an ideal set of complementary experiments
for constraining the dark energy beyond the constant energy density
case \cite{2003PhRvD..67h3505F}, because weak lensing measurement
breaks the degeneracy with the dark energy evolution. 

Earlier work has shown that cosmic shear provides also independent
constraints on $\OmM$ from the measurement of high order statistics
of the convergence \cite{2003astro.ph..2031P,2003AA...397..405B,2002AA...389L..28B,1999AA...342...15V}.
Dark energy modifies mildly this picture. At the level of the quasi-linear
regime, it only affects the three points function of the convergence
field through the projection effect \cite{BB01}. The modifications
are expected to be more important at small scales \cite{1999ApJ...519L...9H}.
One can see from Figure \ref{fig:cfhtls:mod0}, \ref{fig:cfhtls:mod2},
and \ref{fig:cfhtls:mod3} that this additional information is not
necessary, given the degeneracy among $w_{0}$ in the ($w_{0}$, $\OmM$)
space. However, such external constraint could be very helpful to
pin down the degeneracy with $\sigma_{8}$, and consequently to reduce
the degeneracy between $w_{0}$ and $w_{1}$, helping to narrow the
constraint on $w_{1}$.%

\section{conclusion}

We investigated the possibility to constrain the evolution of dark
energy evolution from measurements of the gravitational lensing by
large scale structures. We used the fact that the non-linear growth
rate of structures is significantly affected. This is a consequence
of an earlier influence of dark energy on the expansion of the Universe.
It was found that the cosmic shear effect is an ideal probe of the
evolution of dark energy, in opposition with experiments based on
angular diameter distances like SNIa and cosmic microwave background,
which are better suited to measure the {}``constant'' part of dark
energy equation of state (in a particular parametrization). The degeneracy
with other parameters ($\OmM$, $\sigma_{8}$ and $\Gamma$) restore
a degeneracy between $w_{0}$ and $w_{1}$, but the width of the contours
in that space are slightly affected. Therefore a linear combination
of $w_{0}$ and $w_{1}$ is well measured using weak lensing, even
with a poor knowledge on $\OmM$ and $\sigma_{8}$. 

It is generally believed that the measurement of the dark energy equation
of state parameter as a constant is such a difficult task, that we
should not even dream to measure its evolution. We have shown here,
for a class of models, that the sensitivity of the large scale structures
to a simple evolution parameter $w_{1}$ is as easy (or difficult!)
as $w_{0}$ to measure. Consequently, we found out that a combination
of cosmic shear, SNIa and cosmic microwave background provide orthogonal
constraints of the parameters $w_{0}$, $w_{1}$ and $\OmM$, which
opens great opportunities to probe non-trivial models of dark energy.
For the set of models studied here, we found that these three experiments
over constrain these parameters. 

One should note that the difference in the amplitude of the cosmic
shear signal between model 2 ($w_{0}=-0.8$, $w_{1}=0$) and model
3 ($w_{0}=-0.8$, $w_{1}=0.32$), at scales below $5'$ reaches $10\%$.
This is large compared to the statistical errors of the CFHTLS and
SNAP surveys. However, it is yet within the limits of the Point Spread
Function (PSF) correction and non-linear modeling accuracies \cite{2002AA...393..369V}.
If one wants to measure the dark energy evolution as proposed here,
it is clear that we need to perform ray-tracing simulations for the
class of models we want to investigate, in order to calibrate the
non-linear modeling \cite{2002astro.ph..7664S}. The PSF correction
is an entirely different issue, which is not discussed here, but there
is good hopes to be able to reduce the systematics level by a factor
of 5 to 10 \cite{prep:hoekstra}, which should be enough for our purpose
here. 

Redshift degeneracy was not discussed, but it is not different from
the $\sigma_{8}$ and the $\OmM$ degeneracies. What has been said
for these parameters also applies to the source redshift. In the future,
photometric redshifts will provide accurate source redshift measurements,
as we do not need an accurate redshift for each lensed galaxy. 

\begin{acknowledgements}

The authors acknowledge useful discussions with F. Bernardeau, O. Doré,
B. Fort, M. Joyce, Y. Mellier, D. Pogosyan, R. Scoccimarro and J-P.
Uzan. KB's research is supported by NSF grant PHY-0101738. Our
likelihood computations were run on the NYU Beowulf cluster
supported by NSF grant PHY-0116590.
\end{acknowledgements}

\bibliographystyle{apsrev}
\bibliography{shearQ}

\begin{thebibliography}{52}
\expandafter\ifx\csname natexlab\endcsname\relax\def\natexlab#1{#1}\fi
\expandafter\ifx\csname bibnamefont\endcsname\relax
  \def\bibnamefont#1{#1}\fi
\expandafter\ifx\csname bibfnamefont\endcsname\relax
  \def\bibfnamefont#1{#1}\fi
\expandafter\ifx\csname citenamefont\endcsname\relax
  \def\citenamefont#1{#1}\fi
\expandafter\ifx\csname url\endcsname\relax
  \def\url#1{\texttt{#1}}\fi
\expandafter\ifx\csname urlprefix\endcsname\relax\def\urlprefix{URL }\fi
\providecommand{\bibinfo}[2]{#2}
\providecommand{\eprint}[2][]{\url{#2}}

\bibitem[{\citenamefont{{Perlmutter} et~al.}(1999)\citenamefont{{Perlmutter},
  {Aldering}, {Goldhaber}, {Knop}, {Nugent}, {Castro}, {Deustua}, {Fabbro},
  {Goobar}, {Groom} et~al.}}]{1999ApJ...517..565P}
\bibinfo{author}{\bibfnamefont{S.}~\bibnamefont{{Perlmutter}}},
  \bibinfo{author}{\bibfnamefont{G.}~\bibnamefont{{Aldering}}},
  \bibinfo{author}{\bibfnamefont{G.}~\bibnamefont{{Goldhaber}}},
  \bibinfo{author}{\bibfnamefont{R.~A.} \bibnamefont{{Knop}}},
  \bibinfo{author}{\bibfnamefont{P.}~\bibnamefont{{Nugent}}},
  \bibinfo{author}{\bibfnamefont{P.~G.} \bibnamefont{{Castro}}},
  \bibinfo{author}{\bibfnamefont{S.}~\bibnamefont{{Deustua}}},
  \bibinfo{author}{\bibfnamefont{S.}~\bibnamefont{{Fabbro}}},
  \bibinfo{author}{\bibfnamefont{A.}~\bibnamefont{{Goobar}}},
  \bibinfo{author}{\bibfnamefont{D.~E.} \bibnamefont{{Groom}}},
  \bibnamefont{et~al.}, \bibinfo{journal}{Astrophys. J.}
  \textbf{\bibinfo{volume}{517}}, \bibinfo{pages}{565} (\bibinfo{year}{1999}).

\bibitem[{\citenamefont{{Riess} et~al.}(2001)\citenamefont{{Riess}, {Nugent},
  {Gilliland}, {Schmidt}, {Tonry}, {Dickinson}, {Thompson}, {Budav{\' a}ri},
  {Casertano}, {Evans} et~al.}}]{2001ApJ...560...49R}
\bibinfo{author}{\bibfnamefont{A.~G.} \bibnamefont{{Riess}}},
  \bibinfo{author}{\bibfnamefont{P.~E.} \bibnamefont{{Nugent}}},
  \bibinfo{author}{\bibfnamefont{R.~L.} \bibnamefont{{Gilliland}}},
  \bibinfo{author}{\bibfnamefont{B.~P.} \bibnamefont{{Schmidt}}},
  \bibinfo{author}{\bibfnamefont{J.}~\bibnamefont{{Tonry}}},
  \bibinfo{author}{\bibfnamefont{M.}~\bibnamefont{{Dickinson}}},
  \bibinfo{author}{\bibfnamefont{R.~I.} \bibnamefont{{Thompson}}},
  \bibinfo{author}{\bibfnamefont{T.}~\bibnamefont{{Budav{\' a}ri}}},
  \bibinfo{author}{\bibfnamefont{S.}~\bibnamefont{{Casertano}}},
  \bibinfo{author}{\bibfnamefont{A.~S.} \bibnamefont{{Evans}}},
  \bibnamefont{et~al.}, \bibinfo{journal}{Astrophys. J.}
  \textbf{\bibinfo{volume}{560}}, \bibinfo{pages}{49} (\bibinfo{year}{2001}).

\bibitem[{\citenamefont{{Astier}}(2001)}]{2001PhLB..500....8A}
\bibinfo{author}{\bibfnamefont{P.}~\bibnamefont{{Astier}}},
  \bibinfo{journal}{Physics Letters B} \textbf{\bibinfo{volume}{500}},
  \bibinfo{pages}{8} (\bibinfo{year}{2001}).

\bibitem[{\citenamefont{{Maor} et~al.}(2001)\citenamefont{{Maor}, {Brustein},
  and {Steinhardt}}}]{2001PhRvL..86....6M}
\bibinfo{author}{\bibfnamefont{I.}~\bibnamefont{{Maor}}},
  \bibinfo{author}{\bibfnamefont{R.}~\bibnamefont{{Brustein}}},
  \bibnamefont{and} \bibinfo{author}{\bibfnamefont{P.~J.}
  \bibnamefont{{Steinhardt}}}, \bibinfo{journal}{Physical Review Letters}
  \textbf{\bibinfo{volume}{86}}, \bibinfo{pages}{6} (\bibinfo{year}{2001}).

\bibitem[{\citenamefont{{Goliath} et~al.}(2001)\citenamefont{{Goliath},
  {Amanullah}, {Astier}, {Goobar}, and {Pain}}}]{2001AA...380....6G}
\bibinfo{author}{\bibfnamefont{M.}~\bibnamefont{{Goliath}}},
  \bibinfo{author}{\bibfnamefont{R.}~\bibnamefont{{Amanullah}}},
  \bibinfo{author}{\bibfnamefont{P.}~\bibnamefont{{Astier}}},
  \bibinfo{author}{\bibfnamefont{A.}~\bibnamefont{{Goobar}}}, \bibnamefont{and}
  \bibinfo{author}{\bibfnamefont{R.}~\bibnamefont{{Pain}}},
  \bibinfo{journal}{Astron. \& Astrophys.} \textbf{\bibinfo{volume}{380}},
  \bibinfo{pages}{6} (\bibinfo{year}{2001}).

\bibitem[{\citenamefont{{Brax} et~al.}(2000)\citenamefont{{Brax}, {Martin}, and
  {Riazuelo}}}]{2000PhRvD..62j3505B}
\bibinfo{author}{\bibfnamefont{P.}~\bibnamefont{{Brax}}},
  \bibinfo{author}{\bibfnamefont{J.}~\bibnamefont{{Martin}}}, \bibnamefont{and}
  \bibinfo{author}{\bibfnamefont{A.}~\bibnamefont{{Riazuelo}}},
  \bibinfo{journal}{Phys. Rev. D} \textbf{\bibinfo{volume}{62}},
  \bibinfo{pages}{103505} (\bibinfo{year}{2000}).

\bibitem[{\citenamefont{{Doran} and {Lilley}}(2002)}]{2002MNRAS.330..965D}
\bibinfo{author}{\bibfnamefont{M.}~\bibnamefont{{Doran}}} \bibnamefont{and}
  \bibinfo{author}{\bibfnamefont{M.}~\bibnamefont{{Lilley}}},
  \bibinfo{journal}{Monthly Notices of the RAS} \textbf{\bibinfo{volume}{330}},
  \bibinfo{pages}{965} (\bibinfo{year}{2002}).

\bibitem[{\citenamefont{{Haiman} et~al.}(2001)\citenamefont{{Haiman}, {Mohr},
  and {Holder}}}]{2001ApJ...553..545H}
\bibinfo{author}{\bibfnamefont{Z.}~\bibnamefont{{Haiman}}},
  \bibinfo{author}{\bibfnamefont{J.~J.} \bibnamefont{{Mohr}}},
  \bibnamefont{and} \bibinfo{author}{\bibfnamefont{G.~P.}
  \bibnamefont{{Holder}}}, \bibinfo{journal}{Astrophys. J.}
  \textbf{\bibinfo{volume}{553}}, \bibinfo{pages}{545} (\bibinfo{year}{2001}).

\bibitem[{\citenamefont{{Weller} et~al.}(2002)\citenamefont{{Weller}, {Battye},
  and {Kneissl}}}]{2002PhRvL..88w1301W}
\bibinfo{author}{\bibfnamefont{J.}~\bibnamefont{{Weller}}},
  \bibinfo{author}{\bibfnamefont{R.~A.} \bibnamefont{{Battye}}},
  \bibnamefont{and}
  \bibinfo{author}{\bibfnamefont{R.}~\bibnamefont{{Kneissl}}},
  \bibinfo{journal}{Physical Review Letters} \textbf{\bibinfo{volume}{88}},
  \bibinfo{pages}{231301} (\bibinfo{year}{2002}).

\bibitem[{\citenamefont{{Hu}}(2003)}]{2003PhRvD..67h1304H}
\bibinfo{author}{\bibfnamefont{W.}~\bibnamefont{{Hu}}}, \bibinfo{journal}{Phys.
  Rev. D} \textbf{\bibinfo{volume}{67}}, \bibinfo{pages}{81304}
  (\bibinfo{year}{2003}).

\bibitem[{\citenamefont{{Seljak} et~al.}(2002)\citenamefont{{Seljak},
  {Mandelbaum}, and {McDonald}}}]{2002astro.ph.12343S}
\bibinfo{author}{\bibfnamefont{U.}~\bibnamefont{{Seljak}}},
  \bibinfo{author}{\bibfnamefont{R.}~\bibnamefont{{Mandelbaum}}},
  \bibnamefont{and}
  \bibinfo{author}{\bibfnamefont{P.}~\bibnamefont{{McDonald}}}
  (\bibinfo{year}{2002}), \eprint{astro-ph/0212343}.

\bibitem[{\citenamefont{{Bartelmann}
  et~al.}(2002{\natexlab{a}})\citenamefont{{Bartelmann}, {Meneghetti},
  {Perrotta}, {Baccigalupi}, and {Moscardini}}}]{2002astro.ph.10066B}
\bibinfo{author}{\bibfnamefont{M.}~\bibnamefont{{Bartelmann}}},
  \bibinfo{author}{\bibfnamefont{M.}~\bibnamefont{{Meneghetti}}},
  \bibinfo{author}{\bibfnamefont{F.}~\bibnamefont{{Perrotta}}},
  \bibinfo{author}{\bibfnamefont{C.}~\bibnamefont{{Baccigalupi}}},
  \bibnamefont{and}
  \bibinfo{author}{\bibfnamefont{L.}~\bibnamefont{{Moscardini}}}
  (\bibinfo{year}{2002}{\natexlab{a}}), \eprint{astro-ph/0210066}.

\bibitem[{\citenamefont{{Sereno}}(2002)}]{2002AA...393..757S}
\bibinfo{author}{\bibfnamefont{M.}~\bibnamefont{{Sereno}}},
  \bibinfo{journal}{Astron. \& Astrophys.} \textbf{\bibinfo{volume}{393}},
  \bibinfo{pages}{757} (\bibinfo{year}{2002}).

\bibitem[{\citenamefont{{Benabed} and {Bernardeau}}(2001)}]{BB01}
\bibinfo{author}{\bibfnamefont{K.}~\bibnamefont{{Benabed}}} \bibnamefont{and}
  \bibinfo{author}{\bibfnamefont{F.}~\bibnamefont{{Bernardeau}}},
  \bibinfo{journal}{Phys. Rev. D} \textbf{\bibinfo{volume}{64}},
  \bibinfo{pages}{83501} (\bibinfo{year}{2001}).

\bibitem[{\citenamefont{{Huterer}}(2002)}]{2002PhRvD..65f3001H}
\bibinfo{author}{\bibfnamefont{D.}~\bibnamefont{{Huterer}}},
  \bibinfo{journal}{Phys. Rev. D} \textbf{\bibinfo{volume}{65}},
  \bibinfo{pages}{63001} (\bibinfo{year}{2002}).

\bibitem[{\citenamefont{{Hu}}(2002)}]{2002PhRvD..66h3515H}
\bibinfo{author}{\bibfnamefont{W.}~\bibnamefont{{Hu}}}, \bibinfo{journal}{Phys.
  Rev. D} \textbf{\bibinfo{volume}{66}}, \bibinfo{pages}{83515}
  (\bibinfo{year}{2002}).

\bibitem[{\citenamefont{{Bartelmann}
  et~al.}(2002{\natexlab{b}})\citenamefont{{Bartelmann}, {Perrotta}, and
  {Baccigalupi}}}]{2002AA...396...21B}
\bibinfo{author}{\bibfnamefont{M.}~\bibnamefont{{Bartelmann}}},
  \bibinfo{author}{\bibfnamefont{F.}~\bibnamefont{{Perrotta}}},
  \bibnamefont{and}
  \bibinfo{author}{\bibfnamefont{C.}~\bibnamefont{{Baccigalupi}}},
  \bibinfo{journal}{Astron. \& Astrophys.} \textbf{\bibinfo{volume}{396}},
  \bibinfo{pages}{21} (\bibinfo{year}{2002}{\natexlab{b}}).

\bibitem[{\citenamefont{{Weinberg} and
  {Kamionkowski}}(2003)}]{2003MNRAS.341..251W}
\bibinfo{author}{\bibfnamefont{N.~N.} \bibnamefont{{Weinberg}}}
  \bibnamefont{and}
  \bibinfo{author}{\bibfnamefont{M.}~\bibnamefont{{Kamionkowski}}},
  \bibinfo{journal}{Monthly Notices of the RAS} \textbf{\bibinfo{volume}{341}},
  \bibinfo{pages}{251} (\bibinfo{year}{2003}).

\bibitem[{\citenamefont{{Klypin} et~al.}(2003)\citenamefont{{Klypin},
  {Maccio'}, {Mainini}, and {Bonometto}}}]{2003astro.ph..3304K}
\bibinfo{author}{\bibfnamefont{A.}~\bibnamefont{{Klypin}}},
  \bibinfo{author}{\bibfnamefont{A.~V.} \bibnamefont{{Maccio'}}},
  \bibinfo{author}{\bibfnamefont{R.}~\bibnamefont{{Mainini}}},
  \bibnamefont{and} \bibinfo{author}{\bibfnamefont{S.~A.}
  \bibnamefont{{Bonometto}}} (\bibinfo{year}{2003}), \eprint{astro-ph/0303304}.

\bibitem[{\citenamefont{{Linder} and {Jenkins}}(2003)}]{2003astro.ph..5286L}
\bibinfo{author}{\bibfnamefont{E.~V.} \bibnamefont{{Linder}}} \bibnamefont{and}
  \bibinfo{author}{\bibfnamefont{A.}~\bibnamefont{{Jenkins}}}
  (\bibinfo{year}{2003}), \eprint{astro-ph/0305286}.

\bibitem[{\citenamefont{{Carroll} et~al.}(2003)\citenamefont{{Carroll},
  {Hoffman}, and {Trodden}}}]{2003astro.ph..1273C}
\bibinfo{author}{\bibfnamefont{S.~M.} \bibnamefont{{Carroll}}},
  \bibinfo{author}{\bibfnamefont{M.}~\bibnamefont{{Hoffman}}},
  \bibnamefont{and} \bibinfo{author}{\bibfnamefont{M.}~\bibnamefont{{Trodden}}}
  (\bibinfo{year}{2003}), \eprint{astro-ph/0301273}.

\bibitem[{\citenamefont{Onemli and Woodard}(2002)}]{Onemli:2002hr}
\bibinfo{author}{\bibfnamefont{V.~K.} \bibnamefont{Onemli}} \bibnamefont{and}
  \bibinfo{author}{\bibfnamefont{R.~P.} \bibnamefont{Woodard}},
  \bibinfo{journal}{Class. Quant. Grav.} \textbf{\bibinfo{volume}{19}},
  \bibinfo{pages}{4607} (\bibinfo{year}{2002}), \eprint{gr-qc/0204065}.

\bibitem[{\citenamefont{{Ferreira} and {Joyce}}(1998)}]{1998PhRvD..58b3503F}
\bibinfo{author}{\bibfnamefont{P.~G.} \bibnamefont{{Ferreira}}}
  \bibnamefont{and} \bibinfo{author}{\bibfnamefont{M.}~\bibnamefont{{Joyce}}},
  \bibinfo{journal}{Phys. Rev. D} \textbf{\bibinfo{volume}{58}},
  \bibinfo{pages}{23503} (\bibinfo{year}{1998}).

\bibitem[{\citenamefont{{Malquarti} and {Liddle}}(2002)}]{2002PhRvD..66l3506M}
\bibinfo{author}{\bibfnamefont{M.}~\bibnamefont{{Malquarti}}} \bibnamefont{and}
  \bibinfo{author}{\bibfnamefont{A.~R.} \bibnamefont{{Liddle}}},
  \bibinfo{journal}{Phys. Rev. D} \textbf{\bibinfo{volume}{66}},
  \bibinfo{pages}{123506} (\bibinfo{year}{2002}).

\bibitem[{\citenamefont{{Bartelmann, Matthias} and {Schneider,
  Peter}}(1999)}]{Bartelmann:1999yn}
\bibinfo{author}{\bibnamefont{{Bartelmann, Matthias}}} \bibnamefont{and}
  \bibinfo{author}{\bibnamefont{{Schneider, Peter}}} (\bibinfo{year}{1999}),
  \eprint{astro-ph/9912508}.

\bibitem[{\citenamefont{{Van Waerbeke} and
  {Mellier}}(2003)}]{2003astro.ph..5089V}
\bibinfo{author}{\bibfnamefont{L.}~\bibnamefont{{Van Waerbeke}}}
  \bibnamefont{and} \bibinfo{author}{\bibfnamefont{Y.}~\bibnamefont{{Mellier}}}
  (\bibinfo{year}{2003}), \eprint{astro-ph/0305089}.

\bibitem[{\citenamefont{{Bernardeau} et~al.}(1997)\citenamefont{{Bernardeau},
  {van Waerbeke}, and {Mellier}}}]{1997AA...322....1B}
\bibinfo{author}{\bibfnamefont{F.}~\bibnamefont{{Bernardeau}}},
  \bibinfo{author}{\bibfnamefont{L.}~\bibnamefont{{van Waerbeke}}},
  \bibnamefont{and}
  \bibinfo{author}{\bibfnamefont{Y.}~\bibnamefont{{Mellier}}},
  \bibinfo{journal}{Astron. \& Astrophys.} \textbf{\bibinfo{volume}{322}},
  \bibinfo{pages}{1} (\bibinfo{year}{1997}).

\bibitem[{\citenamefont{{Jain} and {Seljak}}(1997)}]{1997ApJ...484..560J}
\bibinfo{author}{\bibfnamefont{B.}~\bibnamefont{{Jain}}} \bibnamefont{and}
  \bibinfo{author}{\bibfnamefont{U.}~\bibnamefont{{Seljak}}},
  \bibinfo{journal}{Astrophys. J.} \textbf{\bibinfo{volume}{484}},
  \bibinfo{pages}{560} (\bibinfo{year}{1997}).

\bibitem[{\citenamefont{{Schneider} et~al.}(1998)\citenamefont{{Schneider},
  {van Waerbeke}, {Jain}, and {Kruse}}}]{1998MNRAS.296..873S}
\bibinfo{author}{\bibfnamefont{P.}~\bibnamefont{{Schneider}}},
  \bibinfo{author}{\bibfnamefont{L.}~\bibnamefont{{van Waerbeke}}},
  \bibinfo{author}{\bibfnamefont{B.}~\bibnamefont{{Jain}}}, \bibnamefont{and}
  \bibinfo{author}{\bibfnamefont{G.}~\bibnamefont{{Kruse}}},
  \bibinfo{journal}{Monthly Notices of the RAS} \textbf{\bibinfo{volume}{296}},
  \bibinfo{pages}{873} (\bibinfo{year}{1998}).

\bibitem[{\citenamefont{{Peebles}}(1993)}]{1993ppc..book.....P}
\bibinfo{author}{\bibfnamefont{P.~J.~E.} \bibnamefont{{Peebles}}},
  \emph{\bibinfo{title}{{Principles of physical cosmology}}}
  (\bibinfo{publisher}{Princeton Series in Physics, Princeton, NJ: Princeton
  University Press, |c1993}, \bibinfo{year}{1993}).

\bibitem[{\citenamefont{{Hamilton} et~al.}(1991)\citenamefont{{Hamilton},
  {Matthews}, {Kumar}, and {Lu}}}]{1991ApJ...374L...1H}
\bibinfo{author}{\bibfnamefont{A.~J.~S.} \bibnamefont{{Hamilton}}},
  \bibinfo{author}{\bibfnamefont{A.}~\bibnamefont{{Matthews}}},
  \bibinfo{author}{\bibfnamefont{P.}~\bibnamefont{{Kumar}}}, \bibnamefont{and}
  \bibinfo{author}{\bibfnamefont{E.}~\bibnamefont{{Lu}}},
  \bibinfo{journal}{Astrophys. J.} \textbf{\bibinfo{volume}{374}},
  \bibinfo{pages}{L1} (\bibinfo{year}{1991}).

\bibitem[{\citenamefont{{Peacock} and {Dodds}}(1996)}]{1996MNRAS.280L..19P}
\bibinfo{author}{\bibfnamefont{J.~A.} \bibnamefont{{Peacock}}}
  \bibnamefont{and} \bibinfo{author}{\bibfnamefont{S.~J.}
  \bibnamefont{{Dodds}}}, \bibinfo{journal}{Monthly Notices of the RAS}
  \textbf{\bibinfo{volume}{280}}, \bibinfo{pages}{L19} (\bibinfo{year}{1996}).

\bibitem[{\citenamefont{{Cooray} and {Sheth}}(2002)}]{2002PhR...372....1C}
\bibinfo{author}{\bibfnamefont{A.}~\bibnamefont{{Cooray}}} \bibnamefont{and}
  \bibinfo{author}{\bibfnamefont{R.}~\bibnamefont{{Sheth}}},
  \bibinfo{journal}{Phys. Rep.} \textbf{\bibinfo{volume}{372}},
  \bibinfo{pages}{1} (\bibinfo{year}{2002}).

\bibitem[{\citenamefont{{Steinhardt} et~al.}(1999)\citenamefont{{Steinhardt},
  {Wang}, and {Zlatev}}}]{1999PhRvD..59l3504S}
\bibinfo{author}{\bibfnamefont{P.~J.} \bibnamefont{{Steinhardt}}},
  \bibinfo{author}{\bibfnamefont{L.}~\bibnamefont{{Wang}}}, \bibnamefont{and}
  \bibinfo{author}{\bibfnamefont{I.}~\bibnamefont{{Zlatev}}},
  \bibinfo{journal}{Phys. Rev. D} \textbf{\bibinfo{volume}{59}},
  \bibinfo{pages}{123504} (\bibinfo{year}{1999}).

\bibitem[{\citenamefont{{Brax} and {Martin}}(2000)}]{2000PhRvD..61j3502B}
\bibinfo{author}{\bibfnamefont{P.}~\bibnamefont{{Brax}}} \bibnamefont{and}
  \bibinfo{author}{\bibfnamefont{J.}~\bibnamefont{{Martin}}},
  \bibinfo{journal}{Phys. Rev. D} \textbf{\bibinfo{volume}{61}},
  \bibinfo{pages}{103502} (\bibinfo{year}{2000}).

\bibitem[{\citenamefont{{Peebles} and {Ratra}}(1988)}]{1988ApJ...325L..17P}
\bibinfo{author}{\bibfnamefont{P.~J.~E.} \bibnamefont{{Peebles}}}
  \bibnamefont{and} \bibinfo{author}{\bibfnamefont{B.}~\bibnamefont{{Ratra}}},
  \bibinfo{journal}{Astrophys. J.} \textbf{\bibinfo{volume}{325}},
  \bibinfo{pages}{L17} (\bibinfo{year}{1988}).

\bibitem[{\citenamefont{{Alam} et~al.}(2003)\citenamefont{{Alam}, {Sahni},
  {Saini}, and {Starobinsky}}}]{2003astro.ph..3009A}
\bibinfo{author}{\bibfnamefont{U.}~\bibnamefont{{Alam}}},
  \bibinfo{author}{\bibfnamefont{V.}~\bibnamefont{{Sahni}}},
  \bibinfo{author}{\bibfnamefont{T.~D.} \bibnamefont{{Saini}}},
  \bibnamefont{and} \bibinfo{author}{\bibfnamefont{A.~A.}
  \bibnamefont{{Starobinsky}}} (\bibinfo{year}{2003}),
  \eprint{astro-ph/0303009}.

\bibitem[{\citenamefont{{Corasaniti} and
  {Copeland}}(2002)}]{2002astro.ph..5544C}
\bibinfo{author}{\bibfnamefont{P.~S.} \bibnamefont{{Corasaniti}}}
  \bibnamefont{and} \bibinfo{author}{\bibfnamefont{E.~J.}
  \bibnamefont{{Copeland}}}, \bibinfo{journal}{ArXiv Astrophysics e-prints} pp.
  \bibinfo{pages}{5544--+} (\bibinfo{year}{2002}).

\bibitem[{\citenamefont{{Huterer} and {Starkman}}(2002)}]{2002astro.ph..7517H}
\bibinfo{author}{\bibfnamefont{D.}~\bibnamefont{{Huterer}}} \bibnamefont{and}
  \bibinfo{author}{\bibfnamefont{G.}~\bibnamefont{{Starkman}}}
  (\bibinfo{year}{2002}), \eprint{astro-ph/0207517}.

\bibitem[{\citenamefont{{Efstathiou}}(1999)}]{1999MNRAS.310..842E}
\bibinfo{author}{\bibfnamefont{G.}~\bibnamefont{{Efstathiou}}},
  \bibinfo{journal}{Monthly Notices of the RAS} \textbf{\bibinfo{volume}{310}},
  \bibinfo{pages}{842} (\bibinfo{year}{1999}).

\bibitem[{\citenamefont{{Van Waerbeke} et~al.}(2002)\citenamefont{{Van
  Waerbeke}, {Mellier}, {Pell{\' o}}, {Pen}, {McCracken}, and
  {Jain}}}]{2002AA...393..369V}
\bibinfo{author}{\bibfnamefont{L.}~\bibnamefont{{Van Waerbeke}}},
  \bibinfo{author}{\bibfnamefont{Y.}~\bibnamefont{{Mellier}}},
  \bibinfo{author}{\bibfnamefont{R.}~\bibnamefont{{Pell{\' o}}}},
  \bibinfo{author}{\bibfnamefont{U.-L.} \bibnamefont{{Pen}}},
  \bibinfo{author}{\bibfnamefont{H.~J.} \bibnamefont{{McCracken}}},
  \bibnamefont{and} \bibinfo{author}{\bibfnamefont{B.}~\bibnamefont{{Jain}}},
  \bibinfo{journal}{{Astron. \& Astrophys.}} \textbf{\bibinfo{volume}{393}},
  \bibinfo{pages}{369} (\bibinfo{year}{2002}).

\bibitem[{\citenamefont{{Schneider} et~al.}(2002)\citenamefont{{Schneider},
  {van Waerbeke}, {Kilbinger}, and {Mellier}}}]{2002AA...396....1S}
\bibinfo{author}{\bibfnamefont{P.}~\bibnamefont{{Schneider}}},
  \bibinfo{author}{\bibfnamefont{L.}~\bibnamefont{{van Waerbeke}}},
  \bibinfo{author}{\bibfnamefont{M.}~\bibnamefont{{Kilbinger}}},
  \bibnamefont{and}
  \bibinfo{author}{\bibfnamefont{Y.}~\bibnamefont{{Mellier}}},
  \bibinfo{journal}{Astron. \& Astrophys.} \textbf{\bibinfo{volume}{396}},
  \bibinfo{pages}{1} (\bibinfo{year}{2002}).

\bibitem[{\citenamefont{{Bardeen} et~al.}(1986)\citenamefont{{Bardeen}, {Bond},
  {Kaiser}, and {Szalay}}}]{1986ApJ...304...15B}
\bibinfo{author}{\bibfnamefont{J.~M.} \bibnamefont{{Bardeen}}},
  \bibinfo{author}{\bibfnamefont{J.~R.} \bibnamefont{{Bond}}},
  \bibinfo{author}{\bibfnamefont{N.}~\bibnamefont{{Kaiser}}}, \bibnamefont{and}
  \bibinfo{author}{\bibfnamefont{A.~S.} \bibnamefont{{Szalay}}},
  \bibinfo{journal}{Astrophys. J.} \textbf{\bibinfo{volume}{304}},
  \bibinfo{pages}{15} (\bibinfo{year}{1986}).

\bibitem[{\citenamefont{{Spergel} et~al.}(2003)\citenamefont{{Spergel},
  {Verde}, {Peiris}, {Komatsu}, {Nolta}, {Bennett}, {Halpern}, {Hinshaw},
  {Jarosik}, {Kogut} et~al.}}]{2003astro.ph..2209S}
\bibinfo{author}{\bibfnamefont{D.~N.} \bibnamefont{{Spergel}}},
  \bibinfo{author}{\bibfnamefont{L.}~\bibnamefont{{Verde}}},
  \bibinfo{author}{\bibfnamefont{H.~V.} \bibnamefont{{Peiris}}},
  \bibinfo{author}{\bibfnamefont{E.}~\bibnamefont{{Komatsu}}},
  \bibinfo{author}{\bibfnamefont{M.~R.} \bibnamefont{{Nolta}}},
  \bibinfo{author}{\bibfnamefont{C.~L.} \bibnamefont{{Bennett}}},
  \bibinfo{author}{\bibfnamefont{M.}~\bibnamefont{{Halpern}}},
  \bibinfo{author}{\bibfnamefont{G.}~\bibnamefont{{Hinshaw}}},
  \bibinfo{author}{\bibfnamefont{N.}~\bibnamefont{{Jarosik}}},
  \bibinfo{author}{\bibfnamefont{A.}~\bibnamefont{{Kogut}}},
  \bibnamefont{et~al.} (\bibinfo{year}{2003}), \eprint{astro-ph/0202209}.

\bibitem[{\citenamefont{{Frieman} et~al.}(2003)\citenamefont{{Frieman},
  {Huterer}, {Linder}, and {Turner}}}]{2003PhRvD..67h3505F}
\bibinfo{author}{\bibfnamefont{J.~A.} \bibnamefont{{Frieman}}},
  \bibinfo{author}{\bibfnamefont{D.}~\bibnamefont{{Huterer}}},
  \bibinfo{author}{\bibfnamefont{E.~V.} \bibnamefont{{Linder}}},
  \bibnamefont{and} \bibinfo{author}{\bibfnamefont{M.~S.}
  \bibnamefont{{Turner}}}, \bibinfo{journal}{Phys. Rev. D}
  \textbf{\bibinfo{volume}{67}}, \bibinfo{pages}{83505} (\bibinfo{year}{2003}).

\bibitem[{\citenamefont{{Pen} et~al.}(2003)\citenamefont{{Pen}, {Zhang}, {van
  Waerbeke}, {Mellier}, {Zhang}, and {Dubinski}}}]{2003astro.ph..2031P}
\bibinfo{author}{\bibfnamefont{U.}~\bibnamefont{{Pen}}},
  \bibinfo{author}{\bibfnamefont{T.}~\bibnamefont{{Zhang}}},
  \bibinfo{author}{\bibfnamefont{L.}~\bibnamefont{{van Waerbeke}}},
  \bibinfo{author}{\bibfnamefont{Y.}~\bibnamefont{{Mellier}}},
  \bibinfo{author}{\bibfnamefont{P.}~\bibnamefont{{Zhang}}}, \bibnamefont{and}
  \bibinfo{author}{\bibfnamefont{J.}~\bibnamefont{{Dubinski}}}
  (\bibinfo{year}{2003}), \eprint{astro-ph/0302031}.

\bibitem[{\citenamefont{{Bernardeau} et~al.}(2003)\citenamefont{{Bernardeau},
  {van Waerbeke}, and {Mellier}}}]{2003AA...397..405B}
\bibinfo{author}{\bibfnamefont{F.}~\bibnamefont{{Bernardeau}}},
  \bibinfo{author}{\bibfnamefont{L.}~\bibnamefont{{van Waerbeke}}},
  \bibnamefont{and}
  \bibinfo{author}{\bibfnamefont{Y.}~\bibnamefont{{Mellier}}},
  \bibinfo{journal}{Astron. Astropys.} \textbf{\bibinfo{volume}{397}},
  \bibinfo{pages}{405} (\bibinfo{year}{2003}).

\bibitem[{\citenamefont{{Bernardeau} et~al.}(2002)\citenamefont{{Bernardeau},
  {Mellier}, and {van Waerbeke}}}]{2002AA...389L..28B}
\bibinfo{author}{\bibfnamefont{F.}~\bibnamefont{{Bernardeau}}},
  \bibinfo{author}{\bibfnamefont{Y.}~\bibnamefont{{Mellier}}},
  \bibnamefont{and} \bibinfo{author}{\bibfnamefont{L.}~\bibnamefont{{van
  Waerbeke}}}, \bibinfo{journal}{Aston. Astrophys.}
  \textbf{\bibinfo{volume}{389}}, \bibinfo{pages}{L28} (\bibinfo{year}{2002}).

\bibitem[{\citenamefont{{van Waerbeke} et~al.}(1999)\citenamefont{{van
  Waerbeke}, {Bernardeau}, and {Mellier}}}]{1999AA...342...15V}
\bibinfo{author}{\bibfnamefont{L.}~\bibnamefont{{van Waerbeke}}},
  \bibinfo{author}{\bibfnamefont{F.}~\bibnamefont{{Bernardeau}}},
  \bibnamefont{and}
  \bibinfo{author}{\bibfnamefont{Y.}~\bibnamefont{{Mellier}}},
  \bibinfo{journal}{{Astron. \& Astrophys.}} \textbf{\bibinfo{volume}{342}},
  \bibinfo{pages}{15} (\bibinfo{year}{1999}).

\bibitem[{\citenamefont{{Hui}}(1999)}]{1999ApJ...519L...9H}
\bibinfo{author}{\bibfnamefont{L.}~\bibnamefont{{Hui}}},
  \bibinfo{journal}{Astrophys. J.} \textbf{\bibinfo{volume}{519}},
  \bibinfo{pages}{L9} (\bibinfo{year}{1999}).

\bibitem[{\citenamefont{{Smith} et~al.}(2002)\citenamefont{{Smith}, {Peacock},
  {Jenkins}, {White}, {Frenk}, {Pearce}, {Thomas}, {Efstathiou}, {Couchmann},
  and {Consortium}}}]{2002astro.ph..7664S}
\bibinfo{author}{\bibfnamefont{R.~E.} \bibnamefont{{Smith}}},
  \bibinfo{author}{\bibfnamefont{J.~A.} \bibnamefont{{Peacock}}},
  \bibinfo{author}{\bibfnamefont{A.}~\bibnamefont{{Jenkins}}},
  \bibinfo{author}{\bibfnamefont{S.~D.~M.} \bibnamefont{{White}}},
  \bibinfo{author}{\bibfnamefont{C.~S.} \bibnamefont{{Frenk}}},
  \bibinfo{author}{\bibfnamefont{F.~R.} \bibnamefont{{Pearce}}},
  \bibinfo{author}{\bibfnamefont{P.~A.} \bibnamefont{{Thomas}}},
  \bibinfo{author}{\bibfnamefont{G.}~\bibnamefont{{Efstathiou}}},
  \bibinfo{author}{\bibfnamefont{H.~M.~P.} \bibnamefont{{Couchmann}}},
  \bibnamefont{and} \bibinfo{author}{\bibfnamefont{T.~V.}
  \bibnamefont{{Consortium}}} (\bibinfo{year}{2002}),
  \eprint{astro-ph/0207664}.

\bibitem[{\citenamefont{Hoekstra}(in preparation)}]{prep:hoekstra}
\bibinfo{author}{\bibfnamefont{H.}~\bibnamefont{Hoekstra}} (\bibinfo{year}{in
  preparation}).

\end{thebibliography}

\end{document}